\documentclass[twocolumn,preprintnumbers,amsmath,prc,amssymb]{revtex4}
	
	\usepackage[english]{babel}
	\usepackage[utf8]{inputenc}
	
	\usepackage{graphicx}
	\usepackage{dcolumn}
	\usepackage{bm}
	\usepackage{amsmath}
	\usepackage{lipsum}
	
	\usepackage{graphicx}    
	\usepackage{wrapfig}     
	
	\usepackage{nameref}
	
	\usepackage{hyperref}
	\hypersetup{
		colorlinks,
		citecolor=black,
		filecolor=black,
		linkcolor=black,
		urlcolor=black
	}
	
	\usepackage{mathtools}
	\vspace{ 0.9cm}

\begin{document}

\title{The concept of induced surface and curvature tensions and a unified description of the gas of hard discs and hard spheres}

\author{Nazar S. Yakovenko$^1$, Kyrill A. Bugaev$^{1, 2}$,
  Larissa V.  Bravina$^{3}$ and Eugene E. Zabrodin$^{3, 4}$
        }

\affiliation{$^1$Department of Physics, Taras Shevchenko National University of Kyiv, 03022 Kyiv, Ukraine}

\affiliation{$^2$Bogolyubov Institute for Theoretical Physics, Metrologichna str. 14$^B$, Kyiv 03680, Ukraine}

\affiliation{$^3$Department of Physics, University of Oslo, PB 1048 Blindern, N-0316 Oslo, Norway}

\affiliation{$^4$Skobeltzyn Institute of Nuclear Physics, Moscow State University, 119899 Moscow, Russia}

\begin{abstract}
Mathematically rigorous derivation of  the hadron matter  equation of state within  the induced surface and curvature tensions approach is worked out. Such an equation of state allows one to go  beyond the Van der Waals approximation for the interaction potential of hard spheres. The compressibility of a single- and two-component hadron mixtures are found for two- and three-dimensional cases. The obtained  results are compared to  the well known  one- and two-component equations of state of  hard spheres and hard discs. The values of the model parameters which successfully reproduce the above-mentioned equations of state  on different intervals of packing fractions are determined from fitting their compressibility factors. 
It is argued that after some modification the developed approach can be also used to describe the mixtures of  gases of  convex hard particles of different sizes and shapes. \\

\end{abstract}

\maketitle


\section{Introduction}

Elucidation of the effects of dense medium influence on the properties of interaction and, more generally, on the properties of   constituents of the considered system  is an important, but  hard  task of many body theory. In particular, a gradual  transition from the excluded volume regime  which "works" at low densities in a gas of hard spheres to the eigen volume regime which should be used at high densities of hard spheres near the transition to a solid phase is well studied, but the question is how one can generate such a transition  in case of  several different hard-core radii for the relativistic systems in which the number of particle is not conserved. 
From  the famous Isihara-Hadwiger (IH) formula \cite{Isihara1,Hadw1,Isihara2}  for the excluded volume of convex hard  particles $2V^{excl}=2V^{eigen}+S^{eigen}(\overline{R_1}+\overline{R_2})$  one can  easily deduce  that at high densities either the surface term proportional to eigen surface $S^{eigen}$ of particle, or the mean curvature radii $\overline{R_1}$ \& $\overline{R_2}$ of  particles should  disappear from the IH equation, if
one  is able to account for the influence of dense medium.

More specifically,  one can state  that even in the simplest systems like the mixture of gases consisting of several kinds of particles with different hard-core radii, i.e. in a multicomponent case,   there is a problem of how a dense thermal medium modifies the eigen surface and curvature tensions of  the constituents (which can be molecular or nuclear clusters or even quark-gluon plasma bags). 
Although in the research of one-component systems \cite{Simple_Liquids}  of hard spheres and hard discs such effects were not studied yet, from the above example on IH formula it is clear that such effects should be present in these  systems as well. The situation started to improve  with the recent invention of the concept of induced surface tension (IST) \cite{IST1}, which not only allows one to identically rewrite the  hard-core repulsion in terms of volume and surface parts, similarly to the IH equation, but also it allows one to go beyond the second virial coefficient approximation and  to reproduce  the third and fourth virial coefficients of hard spheres \cite{IST2,IST3} using  a single additional parameter compared to the Van der Walls (VdW) equation of state (EoS) for the hard spheres.
	
It turns  out that  the IST concept is very helpful, since it also provides an  essential improvement of modeling the properties of one-component nuclear \cite{IST4} and neutron matter \cite{ISTnm1,ISTnm2} EoS with the minimal number of adjustable parameters. 
Moreover, it helped to resolve  some severe  theoretical problems of multicomponent hadronic \cite{IST2,IST3,IST5} and nuclear \cite{IST1} systems. 
 In contrast to the non-relativistic systems,  in the above mentioned ones the number of particles is not 
	conserved, since  in some cases the typical temperatures are comparable to the mass of lightest hadrons, i.e. pions. 
The conserved 
	quantities are the baryonic, electric and strange charges. Apparently, a transformation of the EoS of multicomponent system with the hard-core interaction  from the canonical ensemble to the grand canonical one is highly nontrivial. Therefore, the IST EoS  \cite{IST1,IST2,IST3,IST4,ISTnm1,ISTnm2,IST5} whose number of equations is always two and it does not depend on the number of different hard-core radii, is a very effective and  convenient  tool to study the multicomponent systems  in the grand canonical ensemble.

However, the IST EoS was derived heuristically  \cite{IST1,IST3} in order to demonstrate the physical source of the surface tension coefficient generated by the interaction  among the constituents. Furthermore, extension of the applicability range of the multicomponent hadronic EoS to the packing fractions  $\eta = \sum_k \rho_k v_k^{eigen} \simeq 0.2-0.22$  \cite{IST2,IST3} (here $\rho_k$ is the particle number density of $k$-th sort of particle and $v_k^{eigen}$ denotes their eigen volume), at which the usual VdW EoS with the  hard-core repulsion is  inapplicable, seems to be a moderate improvement compared to the famous Carnahan-Starling (CS) EoS \cite{CSEoS}  which  for  hard spheres works very well up to $\eta \simeq 0.45$ \cite{Simple_Liquids}. Therefore, in this work we present a mathematically rigorous derivation of the IST EoS
for  any  number of different hard-core radii, generalize it by including the 
equation for the curvature tension, and then we extrapolate it to high densities. 

Since the suggested phenomenological 
approach  is rather general, we apply it not only to the description of the gases of  one- and two-component  hard spheres, but also to the EoS of hard discs analyzed in Refs.  \cite{SantosEoS,BSEoS}.
The auxiliary parameters which are introduced  into  the obtained  equations are determined from fitting the compressibility of the hard spheres and hard discs EoS with one and two hard-core radii.  As we argue, our approach allows one to make an important step towards the development of a unified approach to model the multicomponent  systems with many sorts of constituents and shapes in arbitrary dimensions. 

The work is organized as follows. In Section \ref{sec:derivation} we analyze the excluded volume of the mixture of gases of  Boltzmann particles with different 
hard-core radii and using the self-consistent approach we derive the EoS with the induced surface tension and its generalization which accounts for the curvature tension. Their generalizations which allow one to account for higher virial coefficients of gases of  hard spheres and hard discs are also 
worked out in this Section. Section \ref{sec:IST EoS single-comp} is devoted to the comparison of the IST EoS with the one-component Carnahan-Starling EoS for hard spheres and  the one-component Barrio-Solana EoS for hard discs. A thorough comparison of the IST EoS with the two-component EoS of hard spheres and hard discs is made in Section \ref{sec: IST EoS two-comp}.  A similar comparison of the ISCT EoS is performed in Section \ref{sec:ISCT} in which  we demonstrate that the ISCT EoS is able to accurately describe the whole gaseous phase of two-component mixtures of hard spheres and hard discs. The discussion of the obtained results and some possible applications are given in Section \ref{sec:discussing results}. 

	\section{Derivation of IST and ISCT EoS}	
	\label{sec:derivation}

In the collisions of heavy ions  at high energies  which are intensively studied nowadays  \cite{Review_HIC} the number of particles is not conserved, since the kinetic energy of  particles is comparable with the masses of light hadrons.   Nevertheless,  the dilute hadronic phase  studied 
at the late stage of these collisions  is very successfully described by the hadron resonance gas   model \cite{IST2,IST3} which is just the  multicomponent VdW EoS with hard-core repulsion. Since in the reactions of strongly interacting particles the fundamental charges, namely baryonic, electric and strange charges, are conserved, we are forces to employ the grand canonical ensemble. 

In this work our main target will be a dense gas of hard $D$-dimensional spheres with $D=2$ and $D=3$ with subsequent applications to dense hadronic matter. Therefore, the theoretical scheme will be first written for  3-dimensional spheres with the comments on how to reformulate it to
the case of  2-dimensional spheres (discs). Moreover,  for the illustrative applications  of the general scheme in this paper we will use the typical temperatures
for heavy ion collisions being  in the range $T\in [50; 150]$ MeV and two lightest  hadron species, i.e. $\pi$-mesons (or pions, the lightest meson)  and nucleons (protons and neutrons, the lightest baryons). 

In this section we consider the VdW  approximation for the hard spheres and then generalize it in order to account for the virial coefficients of  higher order. 
	
\vspace*{-2.2mm}
	\subsection{Self-Consistent Treatment of Excluded Volume}	
\vspace*{-2.2mm}
	Consider the mixture of Boltzmann particles with the hard-core radii $\left\lbrace R_n ; n = 1, 2,..., N\right\rbrace $. Note that
	the antiparticles are considered as the independent sorts. 
Using  their second virial coefficients (excluded volumes per particle) $b_{kl}$
 	\begin{equation}
	b_{kl} \equiv \frac{2}{3}\pi(R_k+R_l)^3 ,
	\end{equation}
one can find 
	the excluded volume of all pairs  taken per particle $\overline{V}_{excl}$   as
\vspace*{-2.2mm}
	\begin{equation}
	\label{mean V excl 1}
	\overline{V}_{excl} = \frac{\sum\limits_{k, l=1}^N N_k \frac{2}{3} \pi (R_k+R_l)^3 N_l}{\sum\limits_{k=1}^N N_k} ,
	\end{equation}
	where $N_k$ denotes the number of particles of sort $k$. Opening the brackets in Eq. (\ref{mean V excl 1}) one gets
	\begin{equation}
	\label{mean V excl 2}
	\hspace*{-1.1mm}\overline{V}_{excl} = \frac{2}{3} \pi \frac{\sum\limits_{k, l=1}^N \hspace*{-1.56mm}N_k (R_k^3 +3R_k^2R_l+3R_kR_l^2+R_l^3) N_l}{\sum\limits_{k=1}^N N_k} .
	\end{equation}
	Combining the 1-st  term with the 4-th one  in the numerator on the right-hand side of Eq. (\ref{mean V excl 2}), and the 2-nd term with  the 3-rd one, we can write
	\begin{equation}
	\overline{V}_{excl} = \frac{4}{3}\pi\sum\limits_{l=1}^N N_l R_l^3 + 4\pi \frac{(\sum\limits_{k=1}^N N_k R_k^2)(\sum\limits_{l=1}^N N_lR_l)}{\sum\limits_{k=1}^N N_k} , 
	\end{equation}
	or explicitly
\vspace*{-2.2mm}
	\begin{equation}
	\label{mean V excl 3}
	\overline{V}_{excl} = \sum\limits_{l=1}^N N_lV_l + \sum\limits_{k=1}^N N_k S_k \overline{R} .
	\end{equation}
	Here $V_k = \frac{4}{3}\pi R_k^3$ and $S_k = 4\pi R_k^2$ are, respectively, the eigen volume and eigen surface of $k$-th sort of particles, and  the mean radius $\overline{R}$ is defined as
	\begin{equation}
	\label{mean radius 1}
	\overline{R} = {\sum\limits_{l=1}^N N_lR_l} \biggl/ {\sum\limits_{l=1}^N N_l}.
	\end{equation}
Using this expression below we will self-consistently find   the EoS of such a mixture in the thermodynamic limit.
\vspace*{-2.2mm}
	\subsection{Laplace Transform to Isobaric Ensemble}
\vspace*{-2.2mm}
	Our next assumption is that for an infinite system one can replace $\left\lbrace N_l \right\rbrace$ in (\ref{mean radius 1})
by its statistical mean value $\left\langle N_l\right\rangle $:
	\begin{equation}
	\label{mean radius 2}
	\overline{R} \rightarrow {\sum\limits_{l}\left\langle N_l\right\rangle R_l}\biggl/ {\sum\limits_{l}\left\langle N_l\right\rangle} .
	\end{equation}
	where $\left\langle N_l\right\rangle$ will be calculated self-consistently using the grand canonical ensemble (GCE) partition. 
	In other words, 
	using $\overline{V}_{excl}$ (\ref{mean V excl 3}) with $\overline{R}$ defined by Eq. (\ref{mean radius 2}) one can calculate  the GCE partition and then find from it $\overline{R}$.
	
Introducing the chemical potential  $\mu_k$ for $k$-th sort of  particles, one can write the GCE partition as
	\begin{align}
	& \hspace*{-1.1mm}Z_{GCE}(T,\left\lbrace \mu_k \right\rbrace, V) \equiv \nonumber
	\\
	& \hspace*{-1.1mm}\equiv \sum\limits_{ \left\lbrace N_n \right\rbrace }^\infty  \left[ \prod_{k=1}^{N} \frac{\left[	\phi_k e^{\frac{\mu_k}{T} }(V - \overline{V}_{excl}) \right]^{N_k}}{N_k!} \right] \theta(V - \overline{V}_{excl}) ,
	\label{GCE_partition}
	\end{align}
	where $\phi_k$ is a thermal density of particles
	\begin{equation}
	\label{phi}
	\phi_k = g_k\int\limits\limits_{R^D} \frac{dp^D}{(2\pi\hbar)^D} e^{-\frac{\sqrt{p^2+m_k^2}}{T}} ,
	\end{equation}
	of $k$-th sort of particles with the mass $m_k$ and the degeneracy factor $g_k$.  Eq.  (\ref{phi}) is written  for the dimension $D$. Here $\sqrt{p^2+m_k^2}$ is a relativistic energy of such a particle with the $D$-dimensional vector of momentum $\vec p$, while $T$ denotes the system temperature.\\
\indent
The  Heaviside step function $\theta$ in Eq. (\ref{GCE_partition}) is very important, since it ensures the absence of negative values of  available volume $(V - \overline{V}_{excl})$ and provides
the finite number of all particles for finite volume of the system $V$. Its presence, however, makes hard the evaluation of the GCE partition   (\ref{GCE_partition}).  One can overcome this difficulty by 
	making the Laplace transformation with respect to $V$ to  
	the  isobaric partition (for an appropriate review see  \cite{Reuter08}) defined as 
	\begin{equation}
	\label{isobaric partition}
	\hspace*{-2.2mm}Z_{ISO}(T,\left\lbrace \mu_k \right\rbrace, \lambda) \equiv \int\limits_{0}^{\infty}dV e^{-\lambda V}Z_{GCE}(T,\left\lbrace \mu_k \right\rbrace, V).
	\end{equation}
The latter  can be calculated  by changing the integration variable $dV \rightarrow d(V-\overline{V}_{excl})$. But before one has to define $\left\lbrace \left\langle N_k \right\rangle \right\rbrace$ in the GCE variables. Using the partial $\mu_k$-derivative of the partition  (\ref{GCE_partition}), one can define $\left\langle N_k \right\rangle$ as
	\begin{equation}
	\left\langle N_k \right\rangle \equiv T \frac{\partial}{\partial \mu_k} \ln Z_{GCE}(T,\left\lbrace \mu_k \right\rbrace, V) .
	\end{equation}
	Then Eq. (\ref{mean radius 2})	 for $\overline{R}$  can be written as
	\begin{align}
	\label{eq for mean rad}
	&\overline{R} 
	= \frac{\sum\limits_{k = 1}^{N}R_k\frac{\partial}{\partial \mu_k}\ln Z_{GCE}(T,\left\lbrace \mu_k \right\rbrace, V)}{
	\sum\limits_{k = 1}^{N} \frac{\partial}{\partial \mu_k}\ln Z_{GCE}(T,\left\lbrace \mu_k \right\rbrace, V)} .
	\end{align}
	For the isobaric partition $Z_{ISO}(T,\left\lbrace \mu_k \right\rbrace, \lambda)$ (\ref{isobaric partition}) one gets 
	\begin{align}
	&Z_{ISO}(T,\left\lbrace \mu_k \right\rbrace, \lambda) = \int\limits_{0}^{\infty}dV' e^{-\lambda V'} \times
\nonumber	\\
\label{isobaric partition 2}
	& \sum\limits_{ \left\lbrace N_k \right\rbrace } \prod_{k = 1}^{N} \frac{1}{N_k!} \left[ \phi_k e^{ \frac{\mu_k}{T}} V'   \right]^{N_k} e^{-\lambda \overline{V}_{excl} } \theta(V') \,.\quad
	\end{align}
	Substituting  $\overline{V}_{excl}$ from Eq. (\ref{mean V excl 3}) into Eq. (\ref{isobaric partition 2}), one finds
	\begin{align}
	&Z_{ISO}(T,\left\lbrace \mu_k \right\rbrace, \lambda) = 	\nonumber \\
	&= \int\limits_{0}^{\infty}dV' e^{-\lambda V'} \sum\limits_{ \left\lbrace N_k \right\rbrace } \prod_{k = 1}^{N} \frac{1}{N_k!} \left[ \phi_k e^{ \frac{\mu_k}{T}} V' e^{-\lambda V_k - \lambda \overline{R} S_k}   \right]^{N_k}  \nonumber \\
	&= \int\limits_{0}^{\infty} dV' e^{-\lambda V' + V' \sum\limits_{k = 1}^{N}\phi_k e^{\frac{\mu_k}{T} - \lambda V_k - \lambda \overline{R} S_k} } . 
	\label{isobaric partition 3}
	\end{align}
Performing an integration with respect to variable $dV'$ in Eq. (\ref{isobaric partition 3}), one can get the compact form of the isobaric partition
	\begin{equation}
	Z_{ISO}(T,\left\lbrace \mu_k \right\rbrace, \lambda) = \frac{1}{\lambda - \mathcal{F}(\lambda,T,\left\lbrace \mu_k \right\rbrace)} ,
	\end{equation}			
	where the auxiliary function $\mathcal{F}(\lambda,T,\left\lbrace \mu_k \right\rbrace)$ is as follows
	\begin{equation}
	\mathcal{F}(\lambda,T,\left\lbrace \mu_k \right\rbrace) = \sum\limits_{k = 1}^{N} \phi_k  \exp\left[ \frac{\mu_k - p V_k - \lambda S_k}{T} \right] .
		\end{equation}		
	Now one can find the GCE partition by the inverse Laplace transform 
	\begin{align}
	\label{eq 2.17}
	&Z_{GCE}(T,\left\lbrace \mu_k \right\rbrace, V) = \frac{1}{2\pi i} \int\limits\limits_{\chi - i\infty}^{\chi + i\infty} \hspace*{-2mm} d\lambda e^{\lambda V}Z_{ISO}(T,\left\lbrace \mu_k \right\rbrace, V)
	\nonumber
	\\
	&= \frac{e^{\lambda^{*} V}}{1-\frac{\partial \mathcal{F}}{\partial \lambda}(\lambda,T,\left\lbrace \mu_k \right\rbrace)}\biggl|_{\lambda = \lambda^{*}} ,
	\end{align}
	where the integration contour in the complex $\lambda$-plane is chosen to the right-hand side of the rightmost singularity, i.e. $\chi>\lambda^*$ (see \cite{Reuter08} for more details). From Eq. (\ref{eq 2.17}) in the thermodynamic limit $V \rightarrow \infty$ one finds the system pressure as $p \equiv T \lambda^*$, since $Z_{GCE}(T,\{\mu_k\},V \rightarrow \infty) \sim \exp(pV/T )$ by definition \cite{Huang}. 
	Similarly to the analysis of Ref. \cite{Reuter08}, it can be shown that in the thermodynamic limit $V \rightarrow \infty$  the rightmost singularity of the partition (\ref{eq 2.17}) is the simple pole $\lambda^{*}$ which is solution of the equation
\begin{equation}\label{Eq19}
	\lambda^{*} = \mathcal{F}(\lambda^{*},T,\left\lbrace \mu_k \right\rbrace). 
\end{equation}
 Therefore, one can write equation for pressure $p$ explicitly
	\begin{equation}
	\label{eq for pressure 0}
	p = T \sum\limits_{k = 1}^{N}\phi_k \exp\left[ \frac{\mu_k}{T} - \frac{p V_k}{T} - \frac{p \overline{R} S_k}{T} \right]\,.
	\end{equation}
	The second equation comes from the condition  (\ref{eq for mean rad})
	\begin{equation}
	\overline{R} = \frac{\sum\limits_{k = 1}^{N} R_k \frac{\partial}{\partial \mu_k} \left[ \lambda^{*} V  - \ln (1-\frac{\partial \mathcal{F}}{\partial \lambda^{*}} )\right]  }{\sum\limits_{k = 1}^{N} \frac{\partial}{\partial \mu_k} \left[ \lambda^{*} V  - \ln (1-\frac{\partial \mathcal{F}}{\partial \lambda^{*}} )\right] } \,.
	\end{equation}
	Evidently, for $V \rightarrow \infty$, the second term $ \propto \frac{\ln( )}{V} \rightarrow 0$ disappears and, hence,   one can  arrive to  the following result
	\begin{align}
	\overline{R} &\rightarrow \frac{\sum\limits_{k = 1}^{N}R_k \frac{\partial \lambda^{*}}{\partial \mu_k} }{\sum\limits_{k = 1}^{N} \frac{\partial \lambda^{*}}{\partial \mu_k}} = 
	\label{mean radius 3}
	\frac{\sum\limits_{k = 1}^{N} R_k \phi_k \exp\left[ \frac{\mu_k - p V_k - p \overline{R} S_k}{T} \right] }{\sum\limits_{k = 1}^{N} \phi_k \exp\left[ \frac{\mu_k - p V_k - p \overline{R} S_k}{T} \right]}
	\,.
	\end{align}
	Using Eq.  (\ref{eq for pressure 0}) one can rewrite Eq. (\ref{mean radius 3}) as
	\begin{equation}
	\label{eq for surf tension 0}
	\Sigma \equiv p \overline{R} = T \sum\limits_{k = 1}^{N} R_k \phi_k \exp\left[ \frac{\mu_k - p V_k - \Sigma S_k}{T} \right] \,.
	\end{equation}
From this equation it is clear that quantity $\Sigma  S_k$  is the surface part of free energy of  $k$-th sort of particles which is  induced 
by the hard-core repulsion between the constituents. Therefore, $\Sigma$ is the IST coefficient.\\
\indent
Similarly, one can rewrite  Eq. (\ref{eq for pressure 0}) for pressure as
	\begin{equation}
	\label{eq for pressure}
	p = \sum\limits_{k = 1}^{N} p_k =T \sum\limits_{k = 1}^{N}\phi_k \exp\left[ \frac{\mu_k - p V_k - \Sigma S_k}{T} \right]\,,
	\end{equation}
where the partial pressures $\{ p_k\}$ of each sort of particles are  introduced for convenience.
This is the desired system of equations (\ref{eq for surf tension 0}) and (\ref{eq for pressure}) for the IST coefficient $\Sigma$ and pressure $p$, respectively,  in the VdW  approximation.  
The latter can be realized from the expression for the effective excluded volume of $k$-th sort of particles which we define as 
\begin{equation}
	\label{V_eff_1}
	\widetilde V_k^{eff} \equiv \frac{ p V_k + \Sigma S_k}{p} =   V_k + \overline{R} S_k \,.
\end{equation}
	It is clear that $\widetilde  V_k^{eff}$  is the excluded volume which stays in the exponential functions in Eqs. (\ref{eq for surf tension 0}) and (\ref{eq for pressure}) in front of the system pressure $p$. The right-hand side of Eq. (\ref{V_eff_1}) looks as  the IH equation in which the mean curvature radius of convex particle of $k$-th  sort  is replaced by the mean  radius $\overline{R}$ which is averaged over the statistical  ensemble.\\
\indent 
Our next step is to generalize the above system in order to take into account the higher order virial coefficients of hard D-dimensional  spheres.  The idea of  the  IST approach \cite{IST1,IST2,IST3} is that at high densities the mean radius $\overline{R}$ in Eqs. (\ref{eq for surf tension 0}), (\ref{eq for pressure})  and (\ref{V_eff_1}) should gradually vanish with the increase of pressure. As suggested in \cite{IST1} this can be achieved   by replacing $\Sigma S_k$ in the r.h.s. of Eq.  (\ref{eq for surf tension 0}) as 
	\begin{equation}
	\Sigma S_k \rightarrow \Sigma S_k \alpha_k, \qquad \text{where} \quad \alpha_k >1
	\,,
	\end{equation}
where the auxiliary parameters $\alpha_k$ should be fixed  in such a way that they describe the higher virial coefficients. 
Under this generalization  Eq. (\ref{eq for surf tension 0}) becomes 
	\begin{equation}
	\label{eq for surf tension}
	\hspace*{-1.5mm}\Sigma = \hspace*{-0.55mm}\sum\limits_{k = 1}^{N} \Sigma_k = T \hspace*{-0.55mm}\sum\limits_{k = 1}^{N}\hspace*{-0.55mm} R_k \phi_k \exp \hspace*{-0.55mm}\left[\frac{\mu_k - p V_k -\alpha_k \Sigma S_k}{T} \right] ,
	\end{equation}
where $ \Sigma_k$ denotes the surface tension coefficient of $k$-th sort of particles.\\
\indent
From the one- and multicomponent systems analyzed in Refs. \cite{IST1,IST2,IST3} it is known that 
even for the case of a single auxiliary  parameter $\alpha_k = const = \alpha$ the system 
 (\ref{eq for pressure}) and (\ref{eq for surf tension})  
allows one  to go beyond the VdW  approximation. In this work we  analyze a more general case of two-component systems. \\
\indent
The reason of why  all the parameters $\alpha_k$ must be larger than unity becomes clear from the inspection  of the effective excluded volume 
$\widetilde V_k^{eff}$ of Eq. (\ref{V_eff_1}).  Let us rewrite  Eq. (\ref{V_eff_1}) with the help of generalized equation for the surface tension coefficient (\ref{eq for surf tension}) in terms of partial pressures as
\begin{equation}
	\label{V_eff_2}
	V_k^{eff} \equiv  V_k +  S_k { \sum\limits_{l = 1}^{N}p_l R_l e^{-(\alpha_l-1)S_l \Sigma/T } }\left[ \sum\limits_{l = 1}^{N}p_l \right]^{-1}  \,.
\end{equation}
This equation shows one that for low densities, i.e. for $\Sigma S_k^{max}/T \ll 1$, each exponential in Eq. (\ref{V_eff_2}) can be approximated as $\exp \left[- \frac{(\alpha_l-1)S_l \Sigma}{T} \right] \simeq 1$ and, hence, one recovers Eq. (\ref{V_eff_1}).  However, for high densities
one can easily show  that an opposite inequality $\frac{\Sigma {S_k}}{T} \gg 1$ is valid  for  any $S_k >  0$ and, hence, 
under the condition $\alpha_k > 1$  the mean radius  $\overline{R} \equiv \frac{\Sigma}{p}$ vanishes, i.e. in this limit the excluded 
volume of such particles approaches their eigen volume,  $V_k^{eff} \rightarrow V_k  $. 

A success of this scheme motivates us to extend the IST concept and account for the curvature tension in order to widen  its applicability range. 
But before going into numerics, in the next subsection we briefly demonstrate how the curvature tension emerges for the hard-core repulsion.
Note that in nuclear physics the role of eigen curvature tension in the binding energy of large nuclei  is still under discussion \cite{Brack,Pomorski,Mor:2012,Sagun2019}, since in contrast to the  binding energy generated  by the surface tension it is essentially smaller 
and, hence, requires more sophisticated methods to be reliably determined.  However, it seems that 
both the eigen and  the induced  curvature tensions may be important in the vicinity of the critical point of the liquid-gas phase transition \cite{IST1,Sagun2019}. This fact also motivates us to generalize the IST concept. 

\vspace*{-2.2mm}
	\subsection{Introduction of curvature tension}
\vspace*{-2.2mm}
	Let us now demonstrate how the induced  curvature tension naturally appears from the hard-core repulsion. For this purpose we 
	return 
	to the expression (\ref{mean V excl 2}) for the excluded volume $\overline{V}_{excl}$ per particle. Now we combine 
only the 1-st  term with the 4-th term on the right-hand side of Eq.  (\ref{mean V excl 2}), and do not combine the 2-nd term with  the 3-rd one.
For low densities the contributions coming from the 2-nd and   the 3-rd terms in Eq.  (\ref{mean V excl 2}) are the same, but this is not the case 
for high densities and, hence, it is worth to analyze such an approach in more details. In this way one obtains
	\begin{align}\label{Eq28}
	%
%
	\overline{V}_{excl} &= \sum\limits_{k=1}^N N_k V_k +  2\pi  \sum\limits_{k=1}^N N_k R_k^2 \cdot \sum\limits_{l=1}^N N_l R_l
	\left[ \sum\limits_{l=1}^N  N_l  \right]^{-1} + \nonumber \\
	&+ 2\pi \sum\limits_{k=1}^N N_k R_k \cdot \sum\limits_{l=1}^N N_l R_l^2 \cdot  \left[ \sum\limits_{l=1}^{N} N_l \right]^{-1}.
	\end{align}
With the help of the mean radius squared  $\overline{R^2}$ and   the double perimeter $C_k$  of the $k$-th sort of  particle defined as
	\begin{equation}
	\label{eq 2.27}
	\overline{R^2} \equiv 
	{\sum\limits_{k=1}^{N}N_k R_k^2}\biggl/{\sum\limits_{k=1}^{N}N_k}, ~ C_k \equiv 4 \pi R_k ~ \text{for }
	~ D=3, 
	\end{equation}
one can rewrite Eq. (\ref{Eq28}) as
	\begin{align}\label{Eq30}	
	\overline{V}_{excl} = \sum\limits_{k=1}^{N} N_k V_k + \frac{\overline{R}}{2}\sum\limits_{k=1}^{N} N_k S_k +
	\frac{\overline{R^2}}{2} \sum\limits_{k=1}^{N}N_k C_k .
	\end{align}	
Similarly  to the IST  case of Eq. (\ref{mean radius 2}),  for an infinite system  in Eqs. (\ref{eq 2.27}) and (\ref{Eq30}) we replace each $ N_l $  by its   value $\left\langle N_l\right\rangle$   averaged over the GCE ensemble
	\begin{equation}
	\label{eq 2.28}
	\overline{R^2} \rightarrow 
	{\sum\limits_{k=1}^{N}R_k^2 \left\langle N_k \right\rangle }\biggl/{\sum\limits_{k=1}^{N}\left\langle N_k \right\rangle} .
	\end{equation}
The resulting system will, therefore,  include the 3-rd equation for/with the term $ \overline{R^2}$
	which can be self-consistently  written  in terms of  derivatives of  the GCE partition
	\begin{align}
	&\overline{R^2} = 
\frac{\sum\limits_{k=1}^{N} R_k^2 \frac{\partial}{\partial \mu_k} \ln Z_{GCE}(T,\left\lbrace \mu_k \right\rbrace, V) }{
	\sum\limits_{k=1}^{N} \frac{\partial}{\partial \mu_k} \ln Z_{GCE}(T,\left\lbrace \mu_k \right\rbrace, V) } . 
	\end{align}
Introducing the curvature tension coefficient $K = p \overline{R^2}$ one can cast the resulting system in the VdW approximation as
\begin{align}
\label{Eq33n}
\frac{p}{T} =&   \hspace*{-0.77mm} \sum\limits_{k = 1}^{N} \phi_k \exp \hspace*{-0.55mm}  \left[ \frac{\mu_k}{T} - V_k \frac{p}{T} - S_k \frac{\Sigma}{T} -C_k \frac{K}{T}\right] ,
\\
\label{Eq34n}
\frac{\Sigma}{A T} =&  \hspace*{-0.77mm} \sum\limits_{k = 1}^{N} \hspace*{-0.55mm} R_k \phi_k \exp \hspace*{-0.55mm} \left[ \frac{\mu_k}{T} - V_k  \frac{p}{T} - S_k 
\frac{\Sigma}{T} -C_k \frac{K}{T} \right] ,
\\
\label{Eq35n}
\frac{K}{B T} =&  \hspace*{-0.77mm}  \sum\limits_{k = 1}^{N}\hspace*{-0.55mm}  R_k^2 \phi_k \exp \hspace*{-0.55mm}  \left[ \frac{\mu_k}{T} - V_k  \frac{p}{T} - S_k 
\frac{ \Sigma}{T} -C_k  \frac{K}{T} \right]  , 
\end{align}
where we introduced the auxiliary positive  constants $A>0$ and $B>0$, whose meaning will be discussed in a moment. 
Note that the distribution function of the pressure  (\ref{Eq33n}) contains the terms which correspond  to the surface and curvature 
parts of the free energy.  This is similar to 
the realistic extensions of famous  Fisher droplet model \cite{Fisher} suggested in   \cite{NewFisher1,NewFisher2}  which are used to describe the liquid-gas phase transition. 
However, the principal difference of the derived equations  (\ref{Eq33n})-(\ref{Eq35n}) from  the ones  suggested in  Refs. \cite{NewFisher1,NewFisher2} and used by their followers
is that   the coefficients of surface $\Sigma \equiv A \overline{R} p $ and curvature  $ K\equiv  B \overline{R^2} p $ tensions
are not the  fitting parameters, but are defined by  the system (\ref{Eq33n})-(\ref{Eq35n}). 

	An apparent generalization of  this  system 
	which provides a correct behavior $\overline{R}\rightarrow 0$ and $\overline{R^2}\rightarrow 0$  in the limit of high densities, i.e. vanishing of the surface and curvature tensions in this limit, 
	 is to replace
	$\Sigma S_k \rightarrow \alpha_k \Sigma S_k$ in Eqs.  (\ref{Eq34n}) and (\ref{Eq35n}), and $K C_k \rightarrow \beta_k K C_k$ in Eq. (\ref{Eq35n}), but not in  Eq.  (\ref{Eq34n}):
\begin{align}
\label{Eq36n}
\frac{p}{T} =&   \hspace*{-0.77mm} \sum\limits_{k = 1}^{N} \phi_k \exp \hspace*{-0.55mm}  \left[ \frac{\mu_k}{T} - V_k \frac{p}{T} - S_k \frac{\Sigma}{T} -C_k \frac{K}{T}\right] ,
\\
\label{Eq37n}
\frac{\Sigma}{A T} =&  \hspace*{-0.77mm} \sum\limits_{k = 1}^{N} \hspace*{-0.55mm} R_k \phi_k \exp \hspace*{-0.55mm} \left[ \frac{\mu_k}{T} - V_k  \frac{p}{T} - S_k 
\frac{\alpha_k \Sigma}{T} -C_k \frac{K}{T} \right] ,
\\
\label{Eq38n}
\frac{K}{B T} =&  \hspace*{-0.77mm}  \sum\limits_{k = 1}^{N}\hspace*{-0.55mm}  R_k^2 \phi_k \exp \hspace*{-0.55mm}  \left[ \frac{\mu_k}{T} - V_k  \frac{p}{T} - S_k 
\frac{\alpha_k  \Sigma}{T} -C_k  \frac{\beta_k K}{T} \right]  . 
\end{align}
Note that for such a choice one finds that partial pressure $p_l$, partial surface $\Sigma_l$ and curvature $K_l$ tension coefficient of the particle of sort $l$ are related as 
\begin{eqnarray}
\Sigma_l & =&  A R_l \, p_l \exp\left[ - (\alpha_l -1) \frac{S_l \Sigma}{T} \right] \,,\\ 
K_l & =&  \frac{B}{A} R_l \,\Sigma_l \exp\left[ - (\beta_l -1) \frac{C_l K}{T} \right] \,.
\end{eqnarray}
For $\alpha_l > 1$ and   $\beta_l > 1$ from these relations  in the limit of high pressure $p \rightarrow \infty$ 
 one can immediately deduce that   
 $\Sigma_l \rightarrow \infty$,  $K_l \rightarrow \infty$, but even for $R_l > 0$  one finds that  $\Sigma_l \ll R_l p_l$ and $K_l \ll R_l \Sigma_l$.  Therefore,  in this limit one finds $\Sigma = \sum_l \Sigma_l = A \overline{R} \,  p \ll A \sum_l  R_l p_l$ and 
 $K = \sum_l K_l = B \overline{R^2} \,  p \ll B \sum_l  R_l^2 p_l$, where we accounted for the fact that adding the terms with $R_l=0$ into 
 expressions for $\Sigma$ and $K$  one does not change the inequality. 
 
 Comparing an expression for the average excluded volume (\ref{Eq30}) with Eqs. (\ref{Eq36n})-(\ref{Eq38n}) one finds that 
 Eq. (\ref{Eq30}) corresponds to the choice $A=B=\frac{1}{2}$. However, our experience shows that it is more instructive to consider them as the adjustable parameters. 
For  the hard spheres and hard discs  the coefficients $A$, $B$, $\alpha_k> 1$ and $\beta_k>1$ can  be determined   either from the third,  fourth, fifth   and so on  virial coefficients or from the best description of the compressibility of the system under investigation.


\section{IST EoS for a single-component gas of hard D-dimensional spheres}
\label{sec:IST EoS single-comp}

In this section we analyze  the simplest  IST EoS  for a one-component system.
The equation for system pressure is a one-component version of Eq. (\ref{eq for pressure}), while the explicit  expression 
for the surface tension coefficient   (\ref{eq for surf tension})  is
	\begin{equation}
	\label{Eq41}
	\hspace*{-1.5mm}\Sigma = T   R_1 \phi_1 \exp  \left[\frac{\mu_1 - p V_1 -\alpha_1 \Sigma S_1}{T} \right] .
	\end{equation}
Here   $\alpha_1$ denotes the adjustable  parameter.
			\begin{figure}[tbp]
				\begin{center}
					\hspace*{-5mm}
					\includegraphics[scale=0.5]{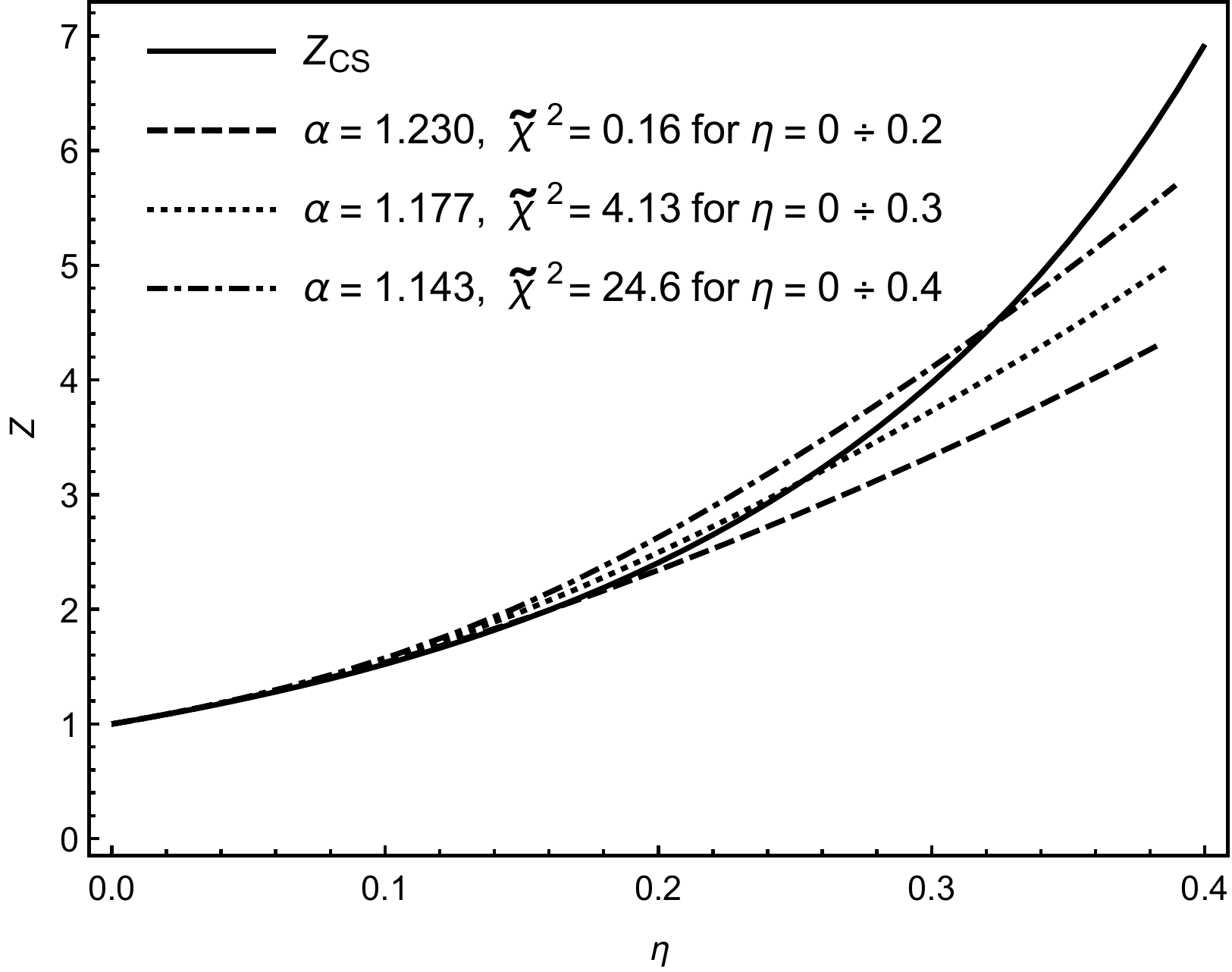}
					\\
					\vspace*{-3mm}
					\caption{Comparison of the compressibility factors of the IST  and CS EoS. The long dashed curve corresponds to the best description of the CS EoS on the interval $\eta\in [0; 0.2]$, the dotted one on the interval $\eta\in [0; 0.3]$ and the dashed-dotted one on the interval $\eta\in [0; 0.4]$.  $\tilde \chi^2 $ is  the mean deviation squared per the interval of fit defined by Eq. (\ref{Eq45}).}
					\label{fig1}
				\end{center}
			\end{figure}

\vspace*{-4.4mm}	

			\subsection{Comparison with Carnahan-Starling EoS}
			\label{sec: IST single comp HS}
			In order to demonstrate the abilities of the IST EoS we compared the compressibility factors $Z$
			\begin{equation}\label{Eq42}
				Z = \frac{p}{\rho T} \,,
		\end{equation}
obtained  from the  IST EoS  and from the CS EoS (\ref{ZCS_scomp}) found for the one-component  gas of hard spheres. 
In Eq. (\ref{Eq42})  $p$ is pressure, while the particle number density $\displaystyle \rho = \frac{\partial p}{\partial \mu_1}$ should be found from 
Eqs. (\ref{eq for pressure}) and (\ref{Eq41}). Some useful formulae for  calculating the particle number density $\rho$  can be found in Appendix \ref{sec:appendix A}. 

The compressibility factor of the CS EoS \cite{CSEoS} is
			\begin{equation}
			\label{ZCS_scomp}
				 Z_{CS} = \frac{1+\eta +\eta^2 -\eta ^3}{(1-\eta )^3} \,, \qquad \eta = \rho V_1 \,,
		\end{equation}
	where $\eta$ is a packing fraction of a considered system and $V_1$ is  the eigen volume of a particle.
The CS EoS very accurately reproduces 12 virial coefficients of the gas of hard spheres  as it is found recently  \cite{MC_for_HS}
from the  Monte Carlo simulations
	\begin{align}\label{Eq44}
		Z_{CS} = &1+ \sum_{n=1}n(n+3) \eta^n = 1 + 4\eta + 10\eta^2 + 18\eta^3 + \nonumber
		\\
		&+  28\eta^4 + 40\eta^5 + 54\eta^6 + 70\eta^7 + ...  \,.
	\end{align}

Using the compressibility factor $Z$ of the IST EoS  we calculated the best-fit values for the parameters $A$ and $\alpha_1$  on different intervals of the packing fraction $\eta$. For a minimization procedure we used the mean deviation squared
 \begin{equation}\label{Eq45}
\tilde \chi^2 \equiv \sum_{n=1}^{N_{max}} \frac{1}{N_{max}} \left[ \frac{Z_{CS }(\eta_n) - Z(\eta_n)}{0.01} \right]^2 \,,
\end{equation}
where  the equidistant mesh was used for the packing fraction $\eta_n$ with typical values for $N_{max} = 20-50 $ depending on the 
length of the studied interval. Such a definition is convenient, since the value $\sqrt{\tilde \chi^2 }$  immediately gives one the mean deviation
in percent, while the quantity $2 \sqrt{\tilde \chi^2 }$ provides a good estimate for the maximal deviation in percent. 
\begin{figure}[tbp]
	\begin{center}
		\hspace*{-10mm}
		\includegraphics[scale=0.5]{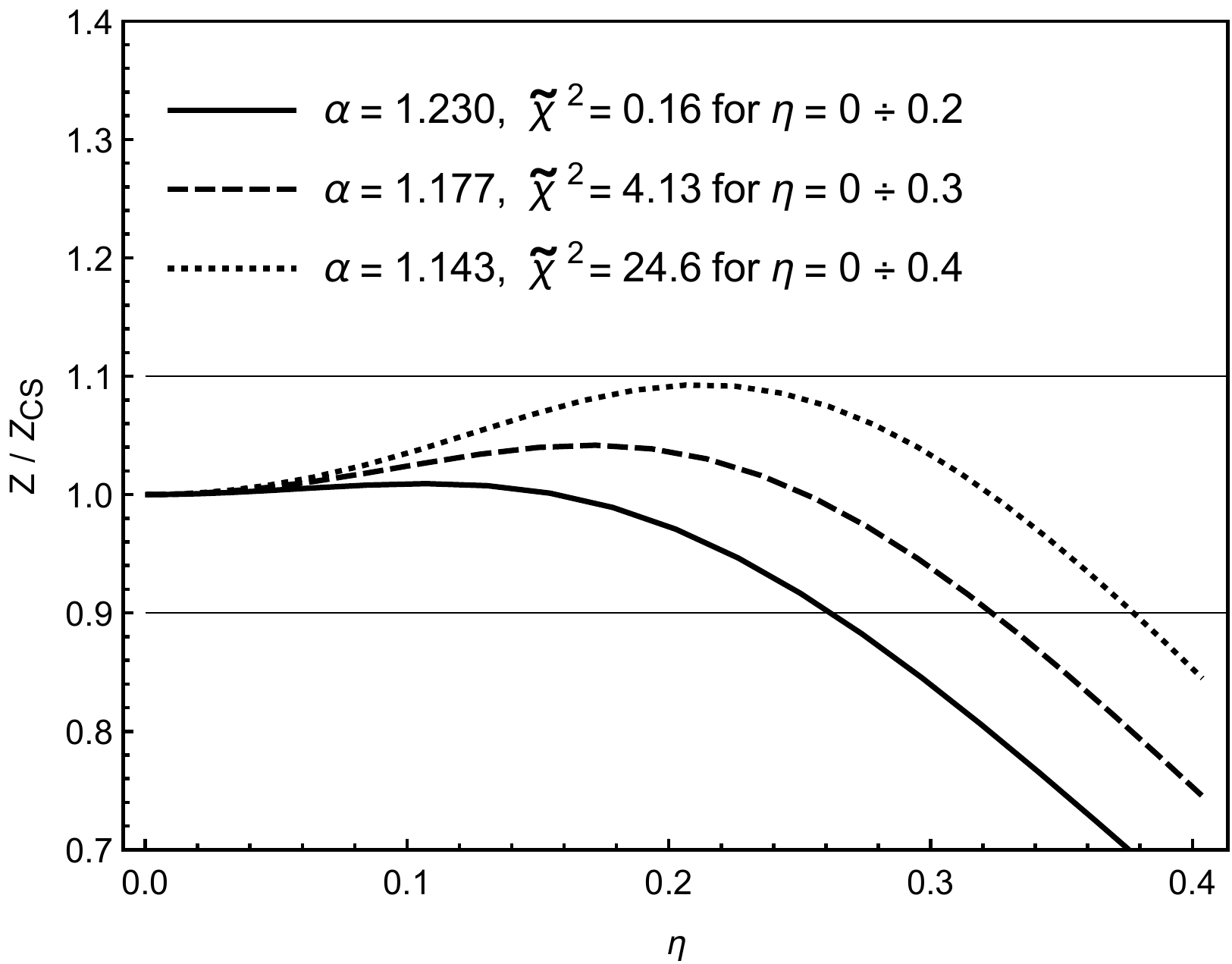}
		\\
		\vspace*{-3mm}
		\caption{ The same as in Fig. \ref{fig1},  but for the ratio of the compressibility factors of IST  and CS  EoS. }
		\label{fig2}
	\end{center}
\end{figure}	
In Figs. \ref{fig1} and \ref{fig2} the compressibility factors of the  IST  and CS EoS are shown for the gas of nucleons, i.e. $g_1 =4$
and $m_1 = 938.9$ MeV for the fixed temperature $T=100$. The variation of the particle number density was achieved by the variation of the nucleon chemical potential $\mu_1$.
From Figs. \ref{fig1} and \ref{fig2} one can see that the IST EoS provides  a reasonable  description with the mean deviation  $\sqrt{\tilde \chi^2 } \simeq 2$ percent  deviation  from  the CS EoS compressibility factor  (\ref{ZCS_scomp}) up to $\eta \simeq 0.28-0.30$ using  two  parameters only. This is an essential improvement of the hard sphere gas description compared to Refs. \cite{IST2,IST3} which  perfectly reproduced the CS EoS  up to $\eta \simeq 0.2$ using a single fitting parameter $\alpha_1$.  Note that the  compressibility factor of  the one-component VdW EoS with the hard core repulsion diverges at  $\eta = 0.25$. 

\vspace*{-4.4mm}
\subsection{Comparison with Barrio-Solana EoS}		
\vspace*{-2.2mm}	

Application of the IST and ISCT EoS to model the properties of the gas of 2-dimensional hard spheres (discs) has not only academic 
interest, but also the practical one. The point is that investigation of the surface deformations of the physical  clusters consisting of the
constituents of finite size (molecular clusters or nuclei) which is necessary to estimate the temperature dependence of  eigen surface tension 
is a typical task of 2-dimensional discs and clusters made of any number of discs \cite{HDM1,HDM2}.  However, the existing model 
of   surface deformations of  physical  clusters developed in Refs. \cite{HDM1,HDM2}  employs the high density approximation 
whereas it seems that   for low temperatures the approximation of excluded (2-dimensional) volume for the surface deformations 
is more  appropriate.  

\begin{figure}[tbp]
	\begin{center}
		\hspace*{-10mm}
		\includegraphics[scale=0.5]{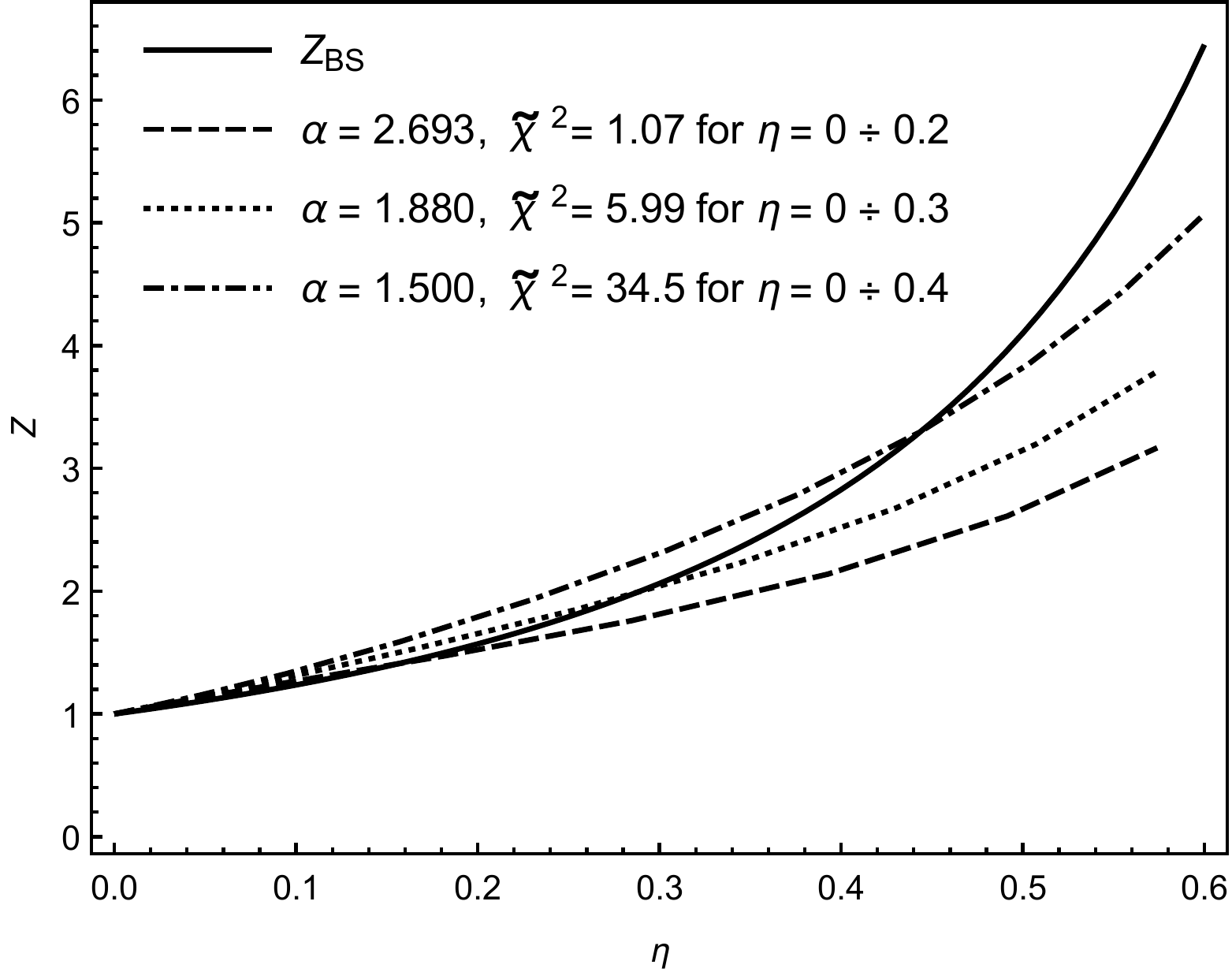}
		\\
		\vspace*{-3mm}
		\caption{Comparison of the compressibility factors of the IST and BS EoS for hard discs. The long-dashed curve corresponds to the best description of the BS EoS on the interval $\eta\in [0; 0.2]$, the dotted one on the interval $\eta\in [0; 0.3]$ and the dashed-dotted one on the interval $\eta\in [0; 0.4]$.  $\tilde \chi^2 $ denotes  the mean deviation squared per the interval of fit.}
		\label{fig3}
	\end{center}
\end{figure}	
 
Bearing in mind such a task  for  future exploration,  in this work we compared  the compressibility factors $Z$ calculated from the  IST EoS for  the gas of hard (hadronic) discs and from the Barrio-Solana (BS) EoS \cite{BSEoS}. The  compressibility factor of the  BS EoS is as follows
\begin{align}
	\label{Z Solana scomp}
		Z_{BS} &= \frac{1+\frac{\eta ^2}{8}-\frac{\eta ^4}{10}}{(1-\eta )^2}
		\,. 
\end{align}
To employ Eqs. (\ref{eq for pressure}) and (\ref{Eq41}) of  the IST EoS  for 2-dimensional case one should 
use the following definitions: $V_1 = \pi R_1^2$ and $S_1 = 2 \pi R_1$ and substitute $D=2$ into Eq. (\ref{phi}) for the thermal density of particles. 

\begin{figure}[tbp]
	\begin{center}
		\includegraphics[scale=0.5]{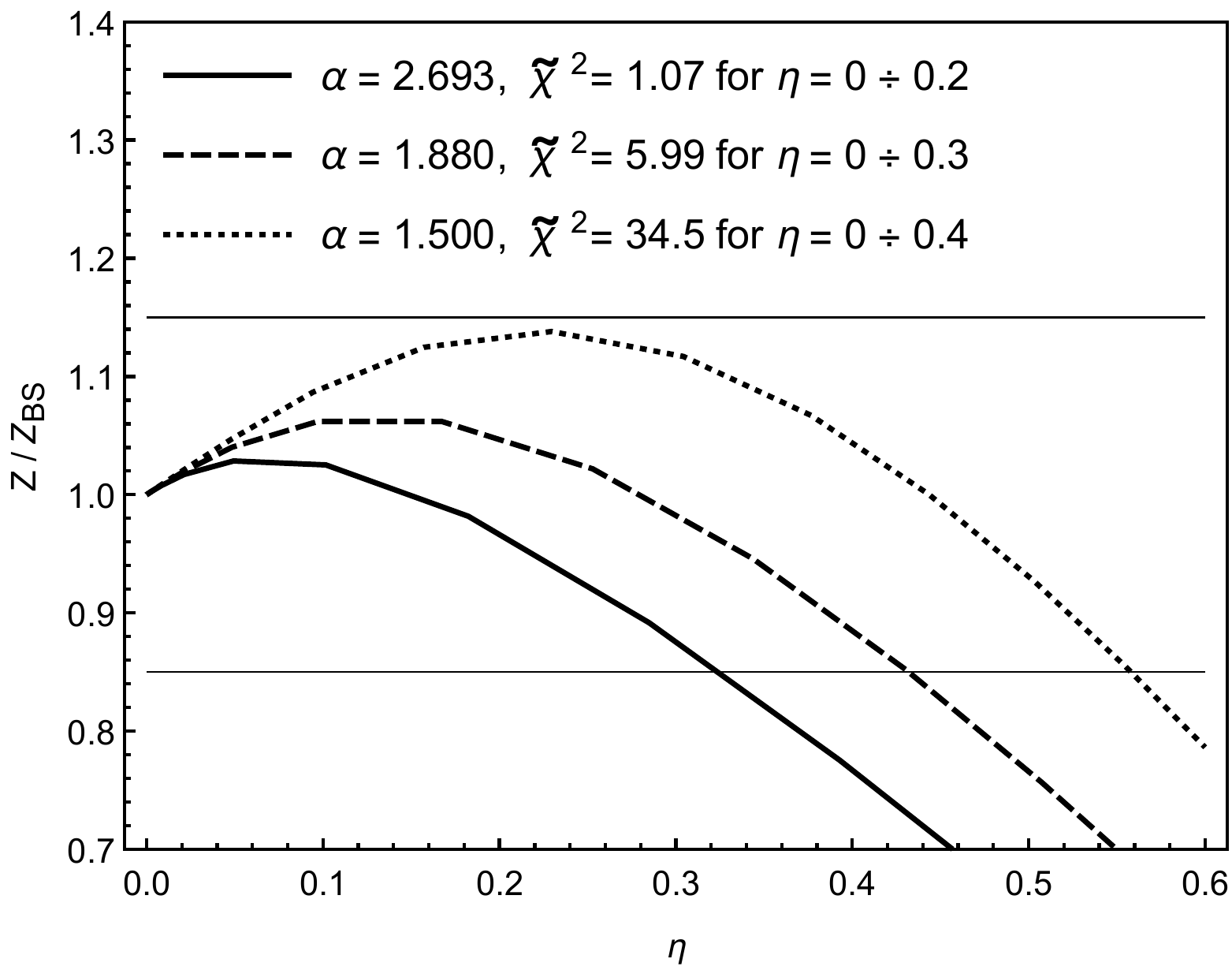}
		\\
		\vspace*{-3mm}
		\caption{Same as in Fig. \ref{fig3},  but for the ratio of compressibility factors of the IST and BS EoS.}
		\label{fig4}
	\end{center}
\end{figure}

Using for the BS EoS the definition of  $\tilde \chi^2$ given by  Eq. (\ref{Eq45}) we found the 
 best-fit parameter  $\alpha_1$  of the  IST EoS for hard discs  on different intervals of the packing fraction $\eta$.
From Figs. \ref{fig3} and Fig. \ref{fig4} one can see that the IST EoS gives a good description  the BS EoS (\ref{Z Solana scomp})
up to $\eta \simeq 0.3$   with $\sqrt{\tilde \chi^2 } \simeq 2.5$ percent   for $\alpha_1 = 1.88$. 

\section{IST EoS for the two-component mixture}
\label{sec: IST EoS two-comp}
\subsection{Comparison with the two-component CS EoS}

Further on we apply  the IST EoS  to description of  the two-component hadron gas of hard spheres and compare it with the multicomponent version of  Carnahan-Starling EoS  known as the Mansoori-Carnahan-Starling-Leland  (MCSL) EoS \cite{MCSL}.
For the  $N$-component mixture  the MCSL  EoS pressure reads as 
\begin{eqnarray}
	\label{ZCS mcomp}
	 \hspace*{-5.5mm}p_{MCSL} &=& \frac{6 T}{\pi } \left[\frac{\xi_0}{1-\xi_3}+\frac{3 \xi_1 \xi_2}{(1-\xi_3)^2}+\frac{3 \xi_2^3-\xi_3 \xi_2^3}{(1-\xi_3)^3}\right],  \quad
	\\
	\text{where} & ~& \xi_n = \frac{\pi}{6} \sum\limits_{k = 1}^{N} \rho_k (2R_k)^n  . \nonumber
\end{eqnarray}
Here we consider  a hadron gas as a mixture of the nucleon-like (with the mass $m_1=938.9$ MeV, the degeneracy factors $g_1=4$ and the particle number density $\rho_1$) and pion-like spheres (with the mass $m_2=138$ MeV, the degeneracy factors $g_2=3$  and the particle number density $\rho_2$). 
For the hard-core radii we used $R_1=0.39$ fm for nucleons and $R_2=0.13$ fm for pions  obtained in Refs. \cite{IST2,IST3}  from fitting the experimental hadron multiplicities to  the hadron resonance gas model.

We calculated the compressibility factors $Z$ for different fixed values of temperature $T$ and baryonic chemical potentials $\mu_1 \neq 0$, $\mu_2=0$ for previously obtained sets of parameters (as in Fig. \ref{fig1}). The quality is basically the same as in Figs. \ref{fig1} and  \ref{fig2}.

\begin{figure}[tbp]
		\centerline{\includegraphics[scale=0.36]{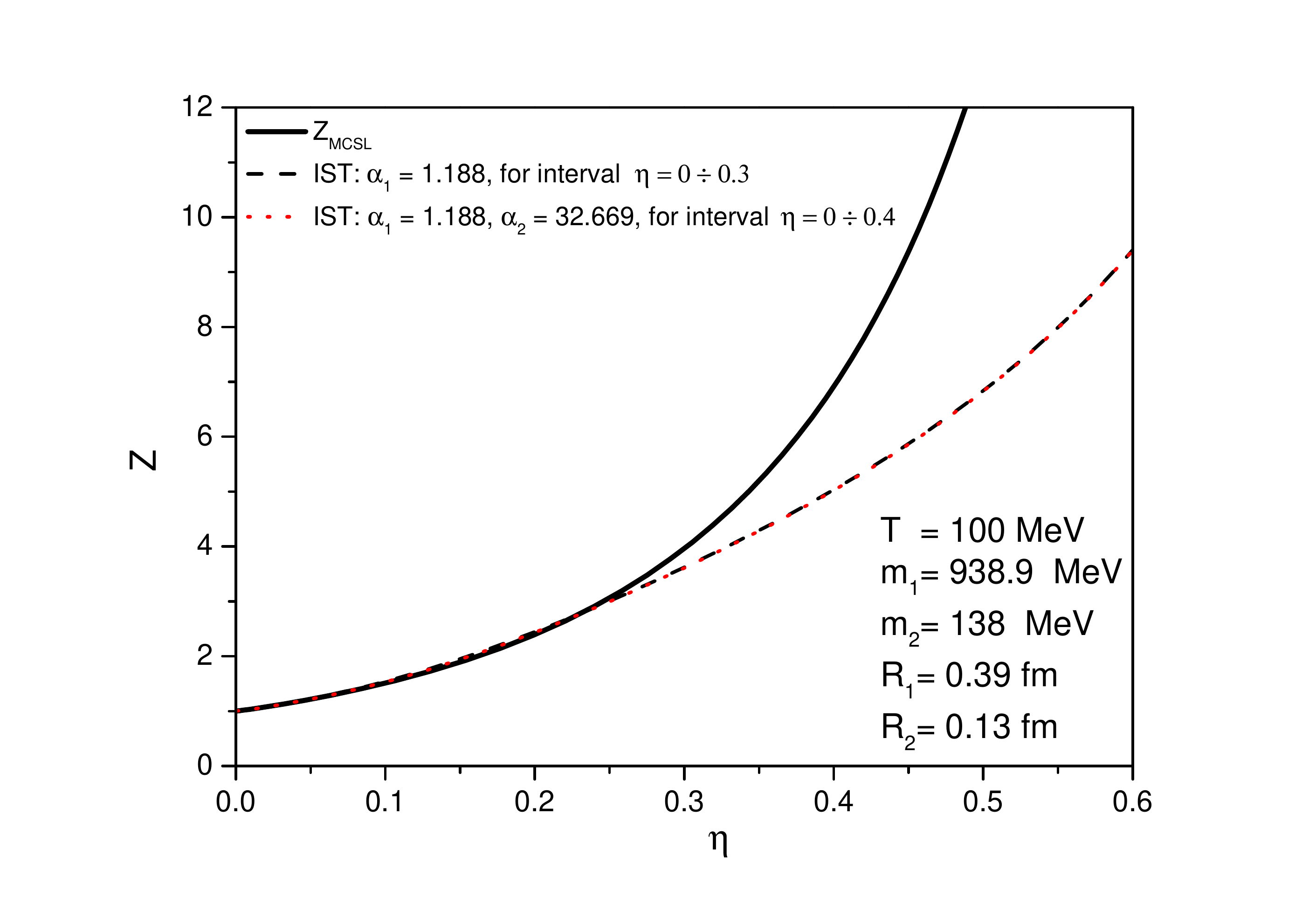}}
		\vspace*{-8mm}
		\centerline{\includegraphics[scale=0.36]{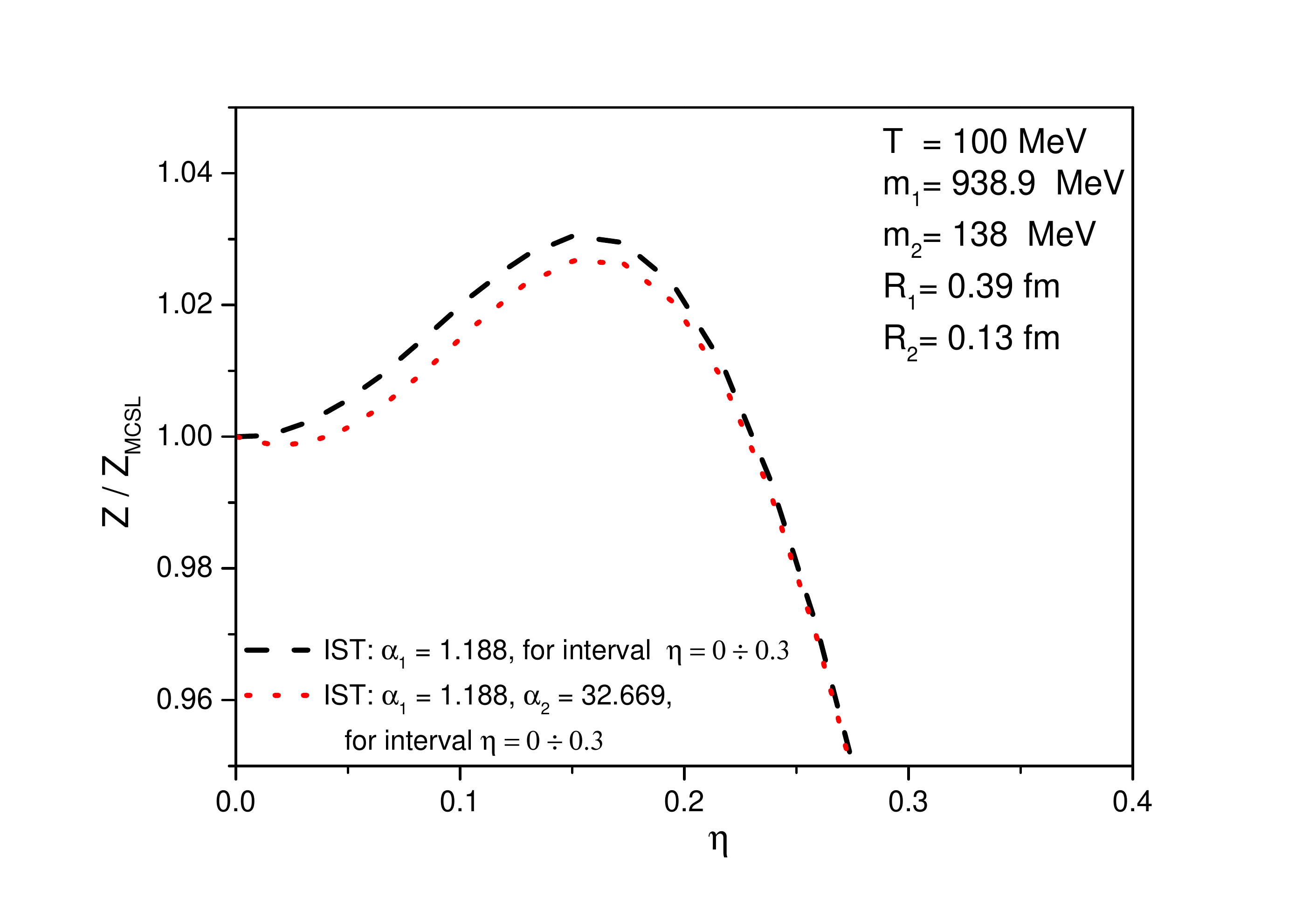}}
		\vspace*{-8mm}
		\caption{ [Color online] {\bf Upper panel.} Compressibility factors $Z$ of the two-component IST EoS and CS EoS (thick solid curve) with the best-fit parameters on interval of packing fraction $\eta \in [0.; 0.3]$. The dashed curve corresponds to the case $\alpha_2=\alpha_1 = 1.188$ with $\tilde \chi^2 \simeq 2.847$, while the  dashed-dotted curve corresponds to the case $\alpha_1 =1.188$, $\alpha_2 =32.669$ and $\tilde \chi^2 \simeq 2.523$. {\bf Lower panel.} Same as in the upper panel, but for the ratios of the compressibility factors $Z/Z_{MCSL}$.}
		\label{fig6}
\end{figure}

The simplest two-component version of the IST coefficient  (\ref{Eq41}) is as follows
	\begin{equation}
	\label{Eq48}
	\hspace*{-1.5mm}\Sigma = T  \sum_{k=1}^2 R_k \phi_k \exp  \left[\frac{\mu_1 - p V_k -\alpha_k \Sigma S_k}{T} \right] .
	\end{equation}
The results for the IST and MCSL EoS are shown in Fig. \ref{fig6}. This figure shows one that introduction of the additional parameter $\alpha_2$ does not improve the quality of $Z_{MCSL}$ fit. 

\vspace*{6.5mm}

	\begin{figure}[thbp]

		\centerline{\includegraphics[scale=0.36]{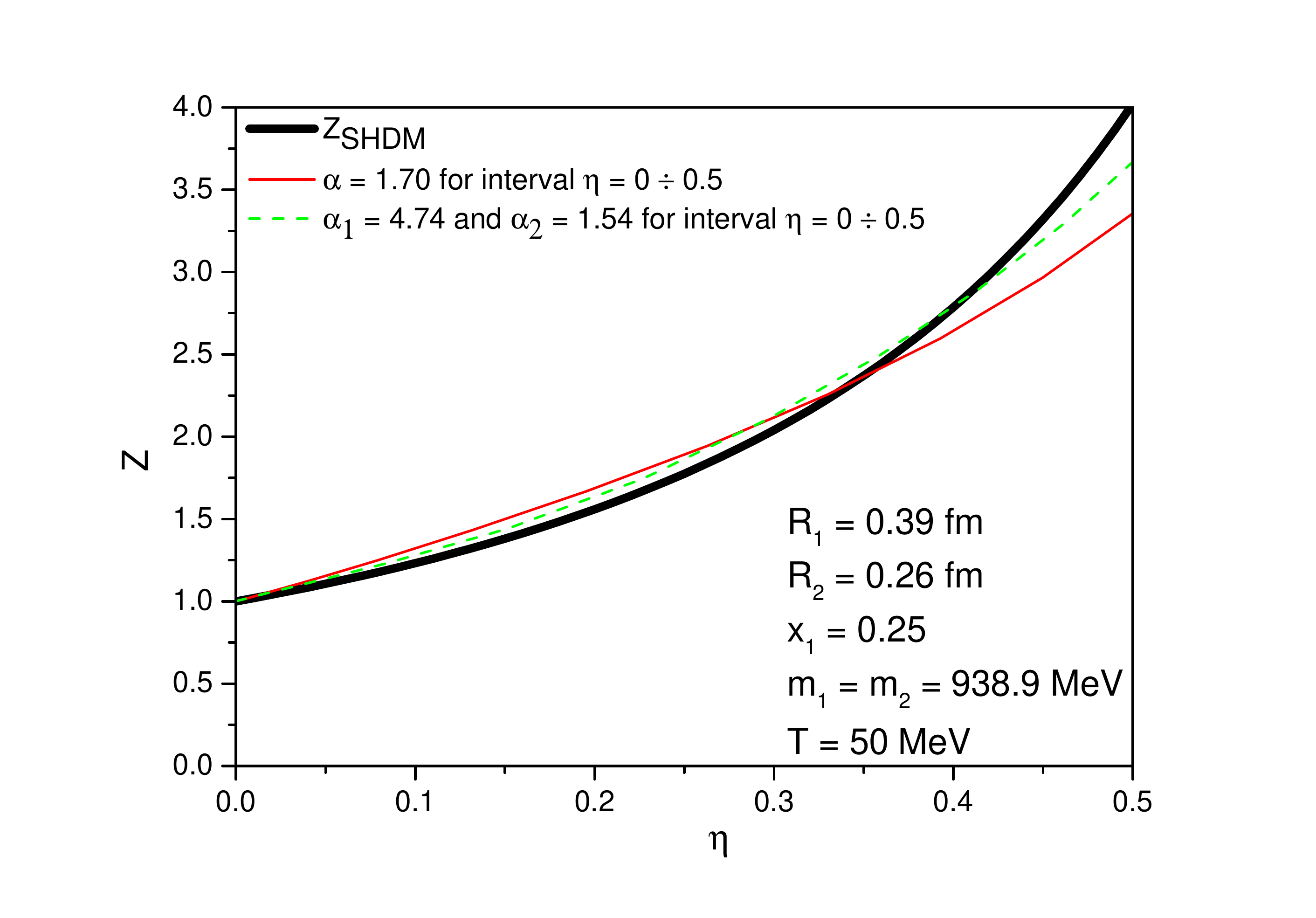}}
		
		\vspace*{-8mm}
		\centerline{\includegraphics[scale=0.36]{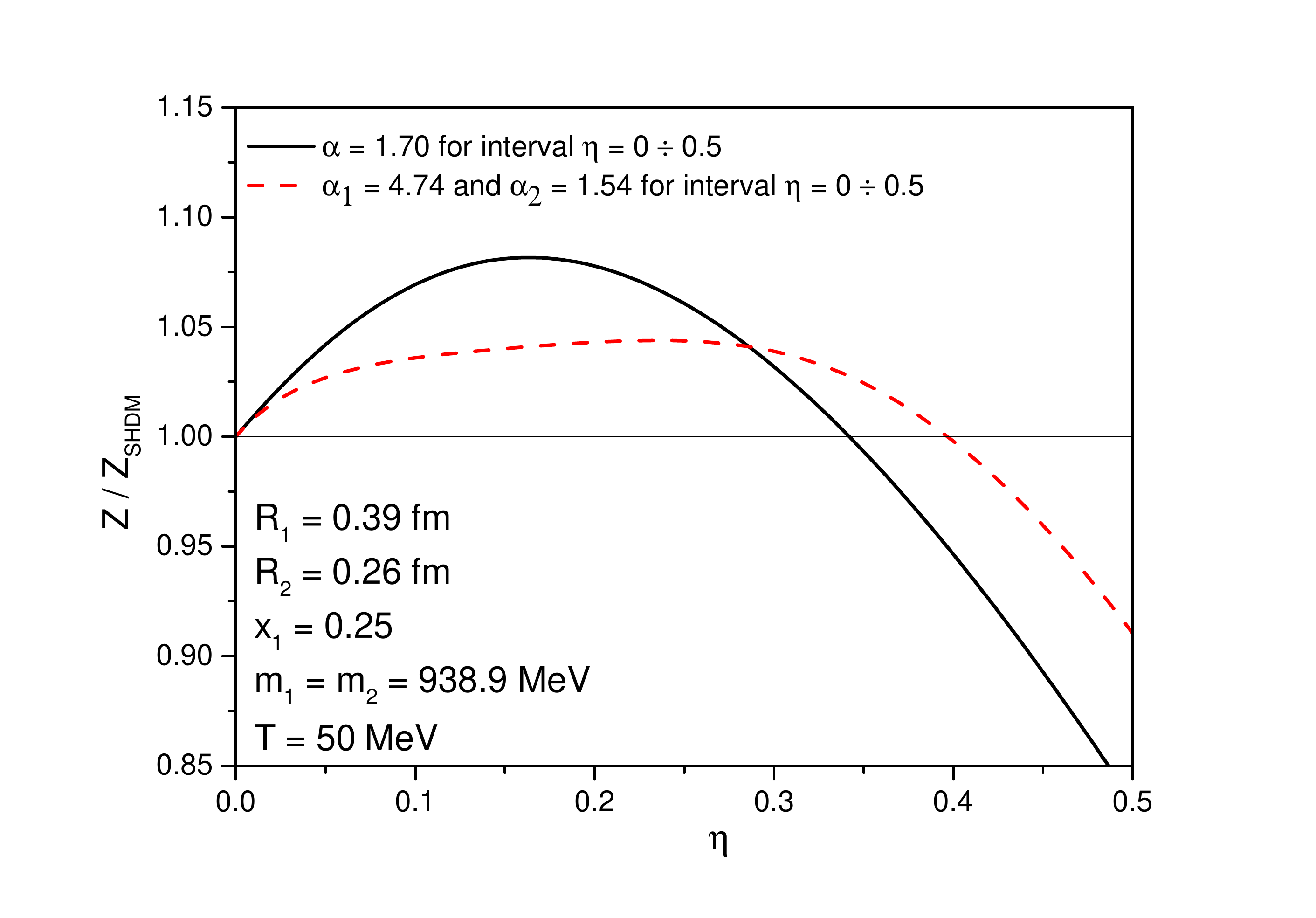}}
		\vspace*{-8mm}
		\caption{{\bf Upper panel.} Compressibility factors $Z$ of the two-component IST EoS and  SHDM EoS (thick solid curve) with the best-fit parameters on interval of packing fraction $\eta \in [0.; 0.5]$. The thin solid curve corresponds to the case $\alpha_2=\alpha_1 = 1.7$ with $\tilde \chi^2 \simeq 14.89$, while the  dashed curve corresponds to the case $\alpha_1 =4.74$, $\alpha_2 =1.54$  and $\tilde \chi^2 \simeq 4.48$. {\bf Lower panel.} Similar to the upper panel, but for the ratios of the compressibility factors $Z/Z_{SHDM}$.}
		\label{fig7}
	\end{figure}

\vspace*{-10.5mm}
\subsection{Comparison with two-component EoS for hard discs}
\label{sec:IST HD twocomp}

	Also we compared the IST EoS with the Santos et al. (SHDM)  EoS \cite{Santos1}  for the compressibility  factor of hard discs two-component mixture 	supplemented by the  Woodcock  EoS \cite{Woodcock} which is used to model  a compressibility of a single component EoS
	(for more details see Ref.  \cite{SantosEoS})
\begin{eqnarray}
\label{Z mult Santos}
	\hspace*{-4.4mm}Z_{SHDM} &=& \frac{(1-\xi)}{1-\eta}+\xi Z_s^W(\eta) , \\
 \label{Woodcock}
 	\hspace*{-4.4mm}Z_s^W(\eta) &=& \frac{1+3\eta / \eta_0}{1 - \eta / \eta_0} + \sum\limits_{n=2}^{6}(b_n \eta_0^{n-1} \hspace*{-1.1mm}-4)\left[\frac{\eta}{\eta_0}\right]^{n-1} 
\hspace*{-1.1mm},~
 \end{eqnarray}
 where the variable 
 \begin{eqnarray}
\xi = \left[ \sum_{j=1}^2 x_j \sigma_j \right]^2 \left[ \sum_{j=1}^2 x_j \sigma_j^2 \right]^{-1} , 
\end{eqnarray}
is expressed in terms of the molar fractions $x_j \equiv \rho_j / (\rho_1+\rho_2)$  and the diameters 
 $\sigma_k = 2R_k$  of hard disc of sort $k$. Apparently, $\rho_k$ denotes the particle number density of the discs of sort $k$. The quantities $b_n$ ($n=2-6$) in Eq. (\ref{Woodcock})  denote the reduced virial coefficients of hard discs  \cite{Vir_Coeff_HD}.

Since   for the two-component case the SHDM  EoS (\ref{Z mult Santos})   requires a fixed concentration for each sort of discs,
we also fix them as $x_1=0.25$ and $x_2=0.75$ for demonstration purpose. 
 In addition, we  implement an extra condition for  the fixed ratio of discs diameters $\sigma_2 / \sigma_1 = 2/3$ in order to make a detailed comparison with   Ref. \cite{SantosEoS}.

For 2-dimensional hadron gas of hard discs mixture of nucleons we used $m_1=m_2=939.8$ MeV and $\sigma_1=0.39$ fm, $\sigma_2=0.26$ fm.  For such input we calculated  the compressibility factor $Z$ for the IST and  SHDM  EoS  and  found the best-fit parameters on various intervals of packing fraction $\eta$. We considered the case of  single $\alpha$ for both constituents and the case of multiple $\alpha_1$, $\alpha_2$. The results for the interval $\eta \in [0.; 0.5]$ are shown in   Fig. \ref{fig7}. As one can see the results found for a single value of $\alpha_2=\alpha_1$ are slightly worse than the ones obtained for the BS EoS shown in Figs.  \ref{fig3} and \ref{fig4} for the case $\alpha_1 =1.88$. For the case of  $\alpha_1 =4.74$ and $\alpha_2 =1.54$ the results  with 
$\tilde \chi^2 \simeq 4.48$ as shown in Fig. \ref{fig7} are comparable to the ones found for the BS EoS.  In principle, the accuracy of few percent achieved  by the IST EoS discussed above would be sufficient for
the most practical applications in high energy nuclear physics, but in our opinion it is a clear signal that the IST concept has to be extended and supplemented by  the curvature term. 

\section{ISCT EoS for hard D-dimensional spheres}
\label{sec:ISCT}

Let us begin this section from a  formal analysis of the IH formula for the excluded volume of a pair of two convex hard  particles. Introducing
now the equivalent sphere radius as $S^{eigen} \equiv 4 \pi R_s^2= S(R_s)$ and doubled perimeters  $C_1 (\overline{R_1}) =4 \pi \overline{R_1}$ and $C_2 (\overline{R_2})
=4 \pi \overline{R_2}$ defined for each mean radius of  curvature $\overline{R_k}$ (with $k=1, 2$) one can exactly rewrite the IH formula
for a pair of  identical convex hard  particles  as 
\begin{eqnarray}\label{Eq52}
2V_{excl}^{IH} & \equiv & \tilde V_1 + a_2 S(\overline{R_1})R_s + (1-\tilde A)C_1 (\overline{R_1}) R_s^2 \nonumber \\
&+ &\tilde V_2
+ a_2 S(\overline{R_2})R_s + (1-\tilde A) C_2 (\overline{R_2})  R_s^2 , \quad 
\end{eqnarray}
where the notations
\begin{eqnarray}\label{Eq53}
\tilde V_1 & = & V^{eigen} -  \tilde B S(\overline{R_1}) R_s (a_1 - a_2) , \\
\tilde V_2 & = & V^{eigen}  + (1+\tilde B) S(\overline{R_1}) R_s (a_1 - a_2) ,\\
\label{Eq54}
a_k & \equiv & \tilde A\frac{R_s}{\overline{R_k}} \quad {\rm with} \quad  k=1, 2 ,
\end{eqnarray}
are used.  In Eqs. (\ref{Eq52}) and (\ref{Eq53})  the arbitrary constants $A$ and $B$ belong to the interval $[0; 1]$. Writing the effective 
excluded volume for particles 1 and 2 from the equation for the pressure of ISCT EoS (\ref{Eq36n})
\begin{eqnarray}\label{Eq56}
&&\hspace*{-5.5mm}2 V_{excl}^{eff}  \equiv   \frac{1}{p} \sum\limits_{k=1}^2 \left[V_k p + S_k \Sigma + C_k K\right] = \\
& &\hspace*{-5.5mm}=  V_1 + A S_1  \overline{R} + B C_1  \overline{R^2} + V_2 + A S_2  \overline{R} + B C_2  \overline{R^2} , ~
\label{Eq57}
\end{eqnarray}
where in deriving Eq.  (\ref{Eq57}) from (\ref{Eq56}) we used the definitions of $\overline{R}$ and $\overline{R^2}$ and Eqs.  (\ref{Eq37n})
and (\ref{Eq38n}).  

Comparing Eqs.  (\ref{Eq52}) and  (\ref{Eq57}) one can conclude that the both expressions have the same structure, if one identifies $R_s  \Leftrightarrow \overline{R}$,  $R_s^2 \Leftrightarrow \overline{R^2}$, $a_2 \Leftrightarrow A$ and $(1-\tilde A)  \Leftrightarrow B$.
Moreover,  this means that for low densities at which the EoS of hard spheres  is defined by the  second virial coefficient 
the ISCT EoS may be used not only for the hard spheres, but also for  the convex hard particles. 
Besides,  this comparison shows that the same excluded volume $V_{excl}^{IH}$ may be reproduced by many sets of parameters
and, hence, one can use this freedom to accurately  account for higher virial coefficients and not to  spoil  the description of the second one. 
Below we demonstrate this for 3- and 2-dimensional  hard spheres. 

\begin{figure}[tbp]

		\centerline{\includegraphics[scale=0.36]{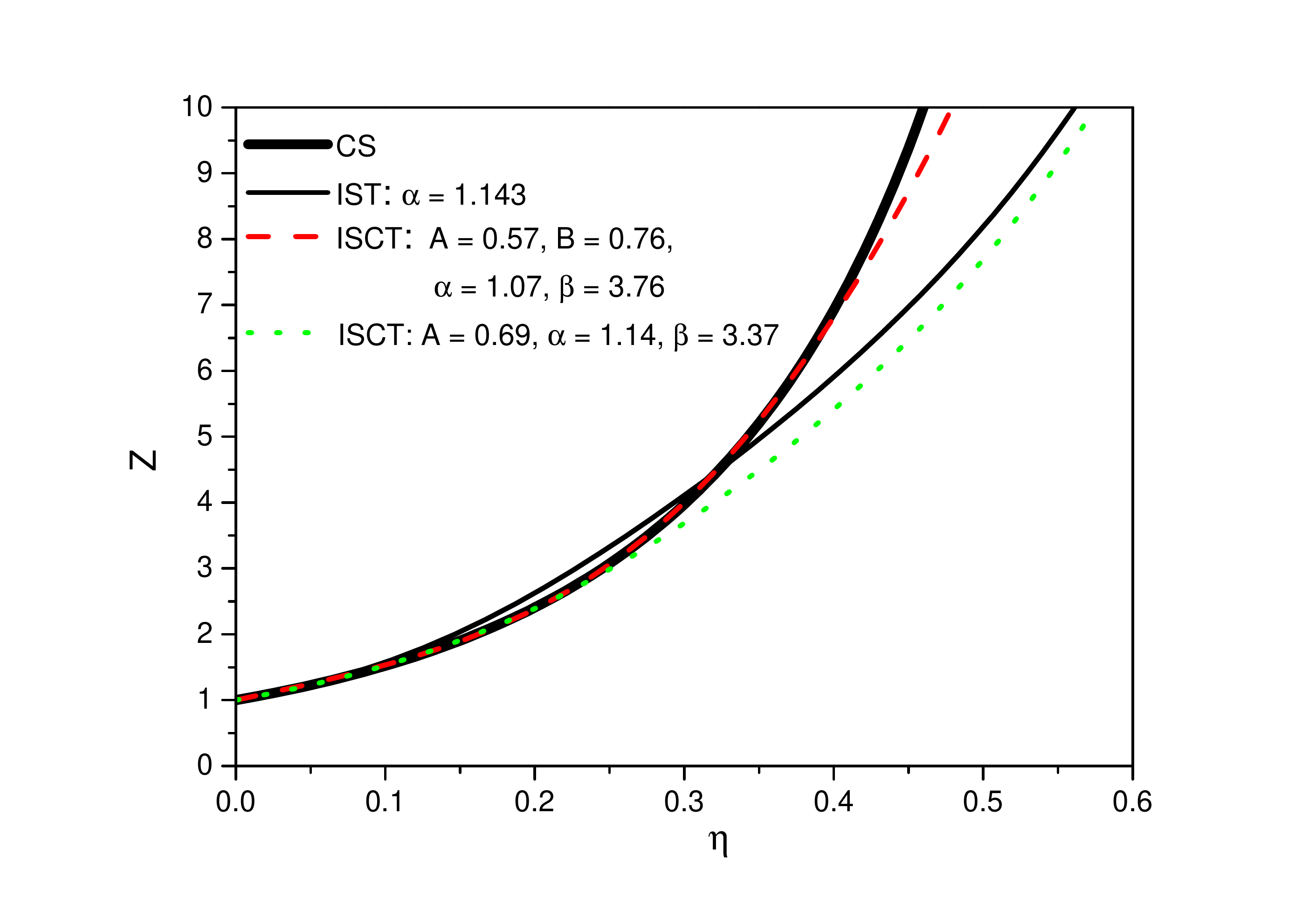}}
		
		\vspace*{-10mm}

			\centerline{\includegraphics[scale=0.36]{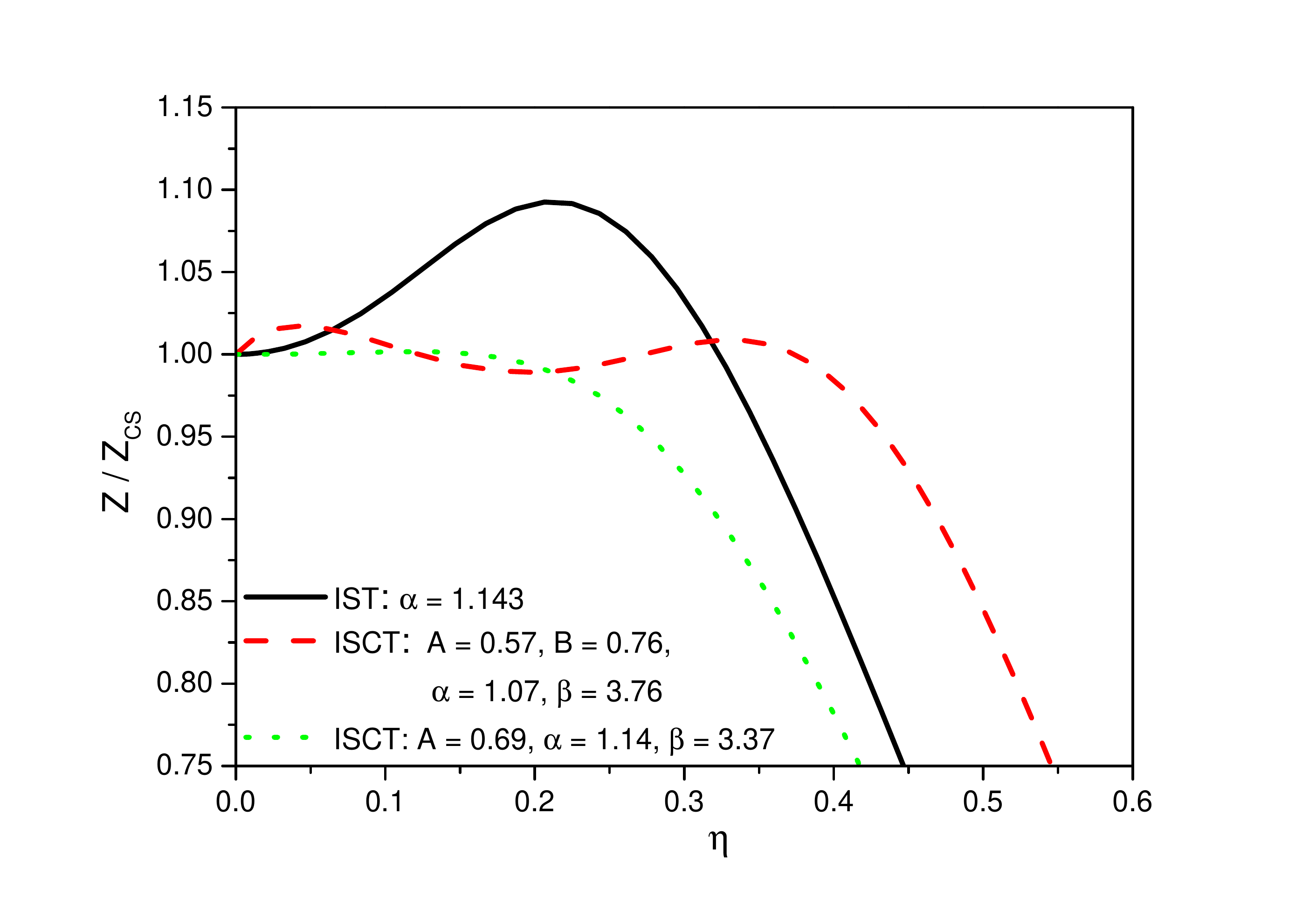}}
	\vspace*{-7.7mm}
	\caption{ [Color online] {\bf Upper panel.} Comparison of the compressibility factors $Z$ of the CS EoS (solid thick curve) with the one-component IST EoS with best-fit parameters on interval of packing fraction $\eta \in [0.; 0.4]$ (from Fig. \ref{fig1},\ref{fig2}) (solid thin curve), the ISCT EoS with best-fit parameters on the same interval with $\tilde \chi^2 \simeq 0.60$ (dashed curve) and  the  ISCT EoS which exactly reproduces the five virial coefficients of the CS EoS (dotted curve). {\bf Lower panel.} Same as in the upper panel, but for the ratios of the compressibility factors $Z/Z_{CS}$.}
	\label{fig8n}
	\end{figure}

For the one-component gas of  hard spheres   we  found that the  set of parameters $A= 0.68$, $B=1-A$,  $\alpha_1= 1.14$ and $\beta=3.37$
the ISCT EoS exactly reproduces the 2-nd, 3-rd, 4-th and 5-th virial coefficients of the CS EoS.  In Fig. \ref{fig8n} the ISCT EoS with such parameters is compared to the IST EoS found  for   the same value of parameter $\alpha_1= 1.14$.  As one can see from Fig. \ref{fig8n} for this set of parameters the ISCT EoS works very well up to the packing fraction $\eta \simeq 0.2$. 
Also from this figure one can see that the ISCT EoS with the best fit parameters $A=0.57$, $B=0.76$, $\alpha_1=1.07$ and $\beta_1=3.76$ is able to provide an essentially  better description of the CS EoS  for the packing fractions $\eta \le 0.45$  with $\tilde \chi^2 = 0.6$. {\it In other words, the ISCT EoS is able to describe the compressibility  of the whole gaseous phase of hard spheres using four parameters only. This is highly nontrivial result, since a comparable quality of description can be achieved by more than  10 virial coefficients of hard spheres.}
The results for a one-component gas of  hard discs are very similar and, hence,  they are not shown. 
	
{
It is interesting that  the one-component ISCT EoS  may help to qualitatively  understand  the reason of why the surface tension of small metallic nano-particles of radius $R_1$  demonstrates the linear dependence on their radius which is justified  theoretically in Refs. \cite{NanoPart1,NanoPart2}. Indeed, considering the dense gas of hard metallic spheres  close to its solid state from Eq.  (\ref{Eq37n})  one can  immediately see that the surface tension coefficient $\Sigma \simeq R_1 F(p, T) $ linearly depends on the radius of hard spheres and on system pressure $p$ and its temperature $T$. Of course, the attraction which exists among the nano-particles studied in Refs. \cite{NanoPart1,NanoPart2} may be important,  but, perhaps,  for small  values of $R_1$ the repulsive effects dominate  and, hence,  the hard-core repulsion  manifests  itself   via the linear dependence of surface tension coefficient on $R_1$. 
}

\begin{figure}[tbp]
	\centerline{\includegraphics[scale=0.36]{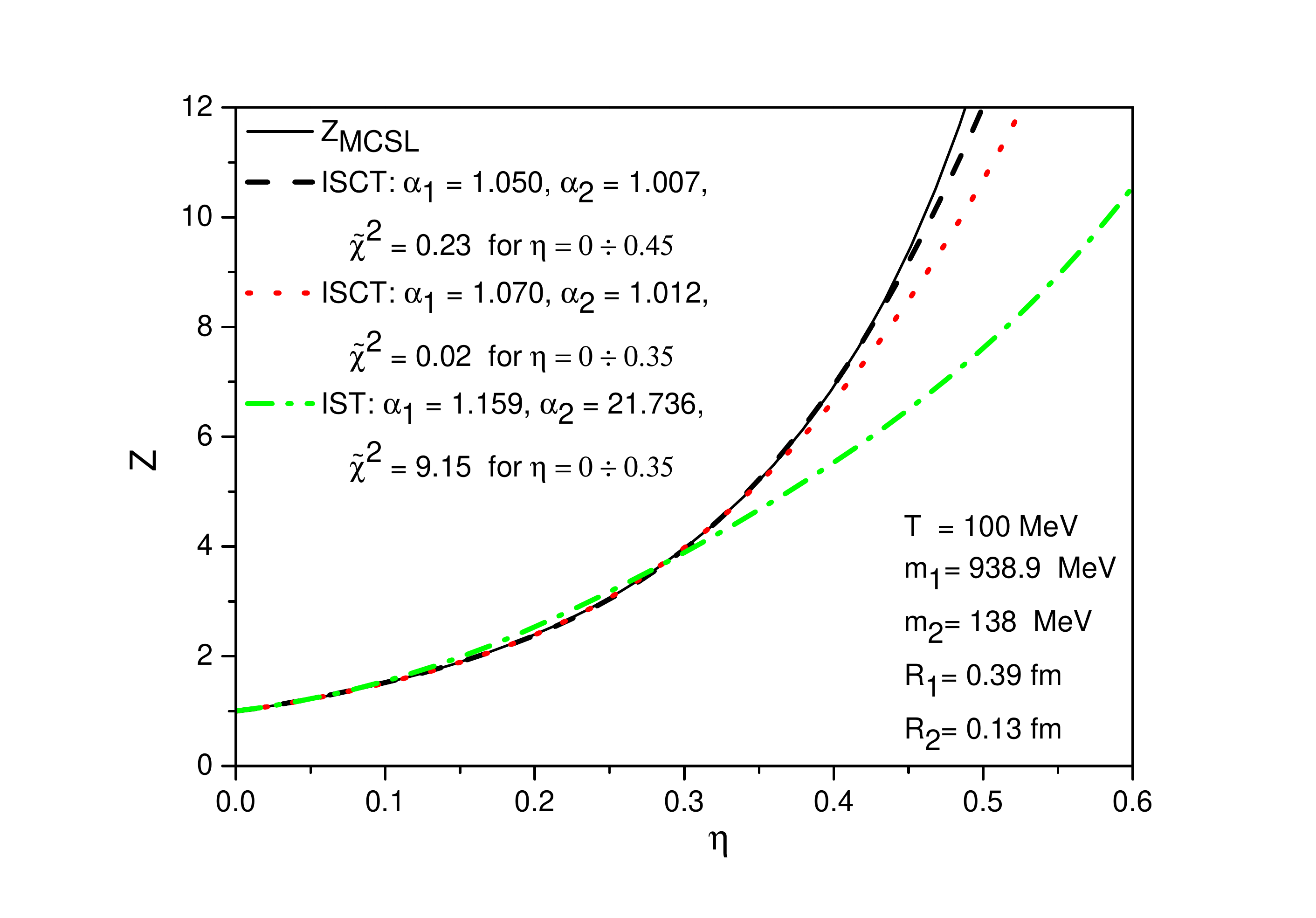}}

	\vspace*{-10mm}

	\centerline{\includegraphics[scale=0.36]{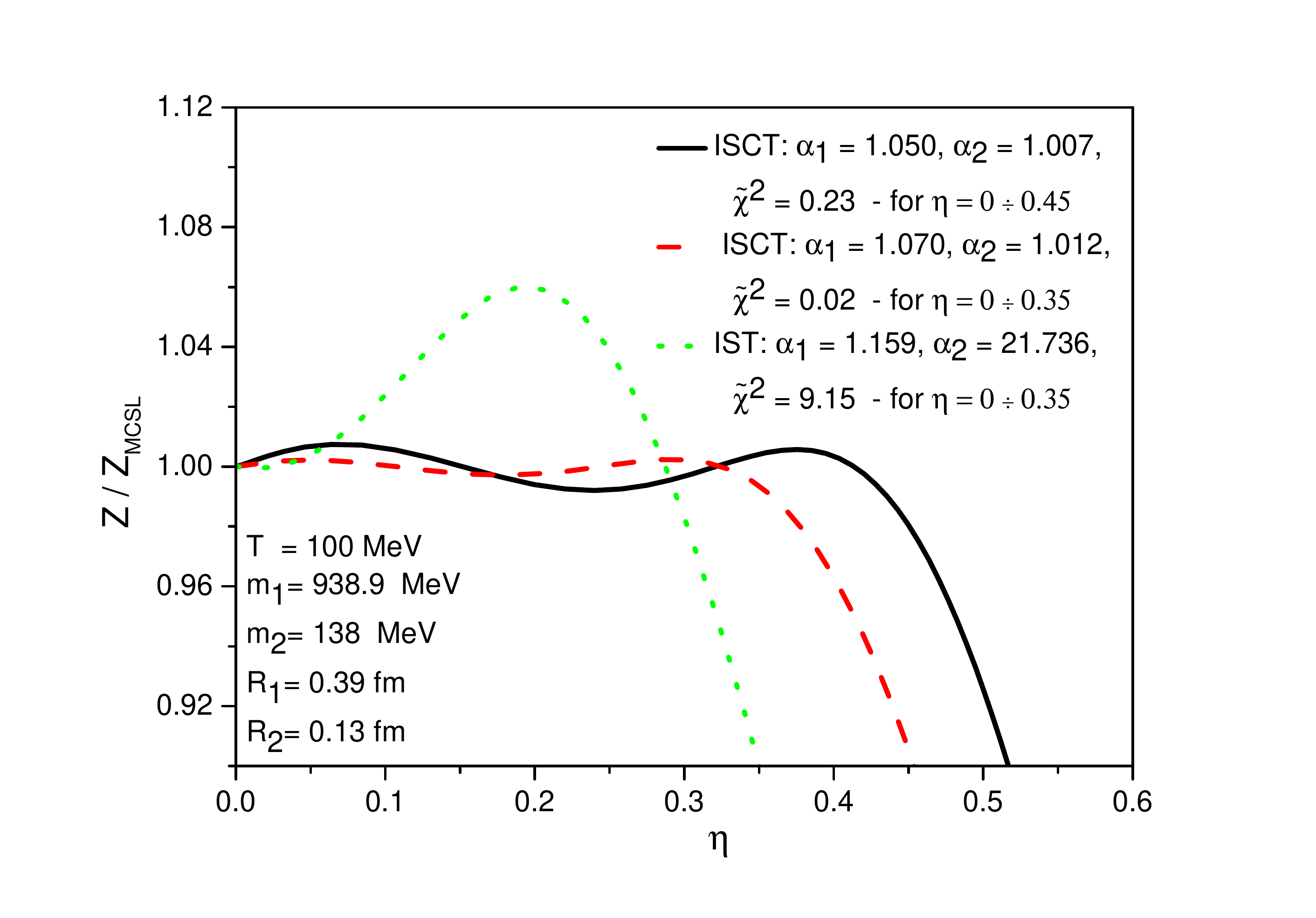}}
	\vspace{-7.7mm}
	\caption{ {\bf Upper panel.} Comparison of the compressibility factors with the best-fit parameters obtained with IST and ISCT EoS for hard spheres gas with multicomponent CS EoS $Z_{MCSL}$ \cite{MCSL}. {\bf Lower panel.} Similar to the upper panel, but for the ratios of the compressibility factors $Z/Z_{MCSL}$ (All parameters are given  in Table. \ref{HS_table}).}
	\label{ISCT HS twocomp}
\end{figure}

Now we turn to the analysis of two-component case.  The ISCT EoS results obtained  for the  mixture of nucleons and pions studied in preceding sections are shown in Fig. 
\ref{ISCT HS twocomp}. This figure  shows one that the two-component CS EoS (\ref{ZCS mcomp}) is very well 
reproduced  by the ISCT EoS on intervals of packing fraction  $\eta = 0.4-0.45$, i.e. almost for an entire gaseous phase of hard spheres. 
The best fit parameters of ISCT EoS which correspond to Fig. \ref{ISCT HS twocomp} are given in Table \ref{HS_table}.
Note that only  with  two additional parameters compared  to the one-component case we obtained even better  quality of the fit. 
\begin{table}[hb]
	\centering
	\caption{The best fit parameters of the ISCT EoS which reproduce the MCSL EoS on two intervals. The upper row corresponds to 
	to the fitting interval $\eta \in [0.;  0.45]$, while the  lower one provides an excellent description at the interval $\eta \in [0.;  0.35]$.}
	\label{HS_table}
	\begin{tabular}{l|l|l|l|l|l|l|l}
		$\alpha_1$ & $\alpha_2$ & $\beta_1$ & $\beta_2$ & $A$ & $B$ & $\tilde \chi^2$ & $\eta$ \\ 
		\hline
		\hline
		1.050 & 1.007 & 2.084 & 1.862 & 0.442 & 0.630 & 0.23 & 0 - 0.45 \\ \hline
		1.070 & 1.012 & 2.428 & 2.372 & 0.520 & 0.512 & 0.02 & 0 - 0.35 \\ 
		
	\end{tabular}
\end{table}

\begin{figure}[htbp]
		\centerline{\includegraphics[scale=0.36]{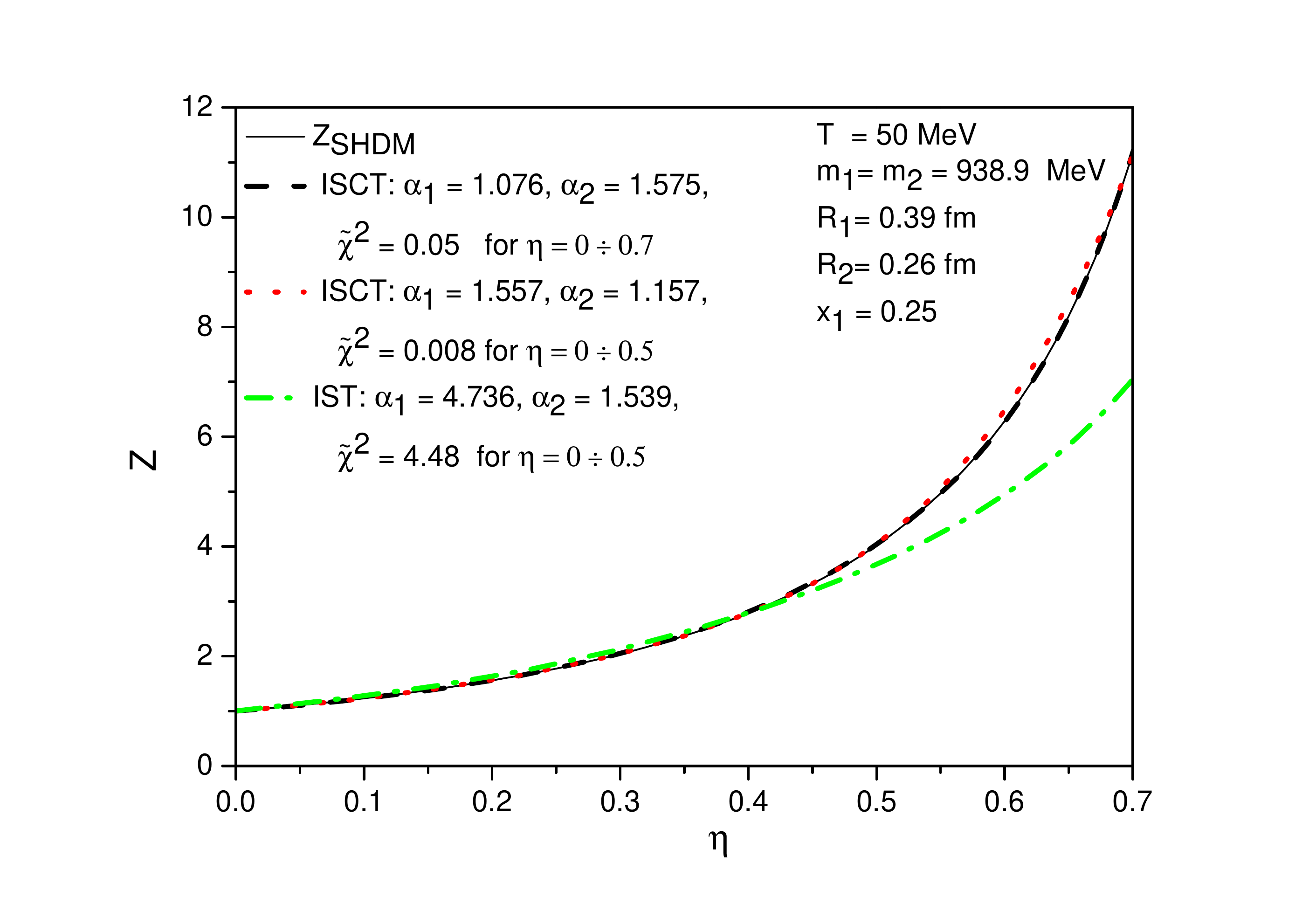}}
		\vspace*{-10mm}

		\centerline{\includegraphics[scale=0.36]{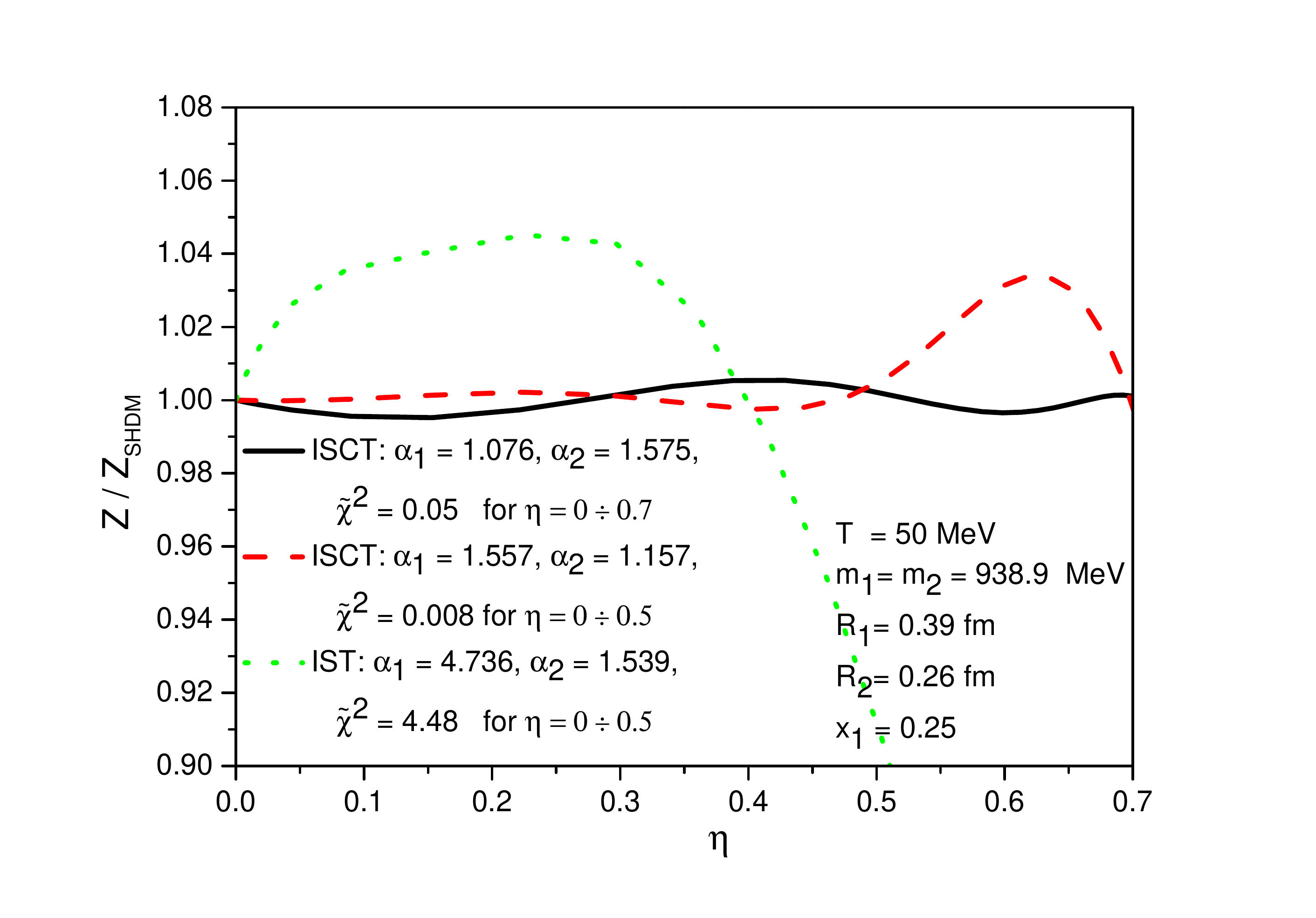}}
	\vspace*{-7.7mm}
	\caption{ {\bf Upper panel.} Comparison of the compressibility factors with the best-fit parameters obtained with IST and ISCT EoS for hard discs mixture with multicomponent SHDM  EoS $Z_{SHDM}$ \cite{SantosEoS}. {\bf Lower panel.} Same as in the upper panel, but for the ratios of the compressibility factors $Z/Z_{SHDM}$ (All parameters are given  in Table. \ref{HD_table}).}
	\label{ISCT HD twocomp}
	\end{figure}

Fig. \ref{ISCT HD twocomp} shows one  the ratio of ISCT EoS compressibility factor  $Z$ for two-component hadron gas of hard discs  to the one  of the   SHDM EoS (\ref{Z mult Santos}) with the best-fit parameters up to $\eta = 0.7$ for the  case   analyzed  in Section (\ref{sec:IST HD twocomp}). In 2-dimensional case the ISCT EoS gives even more notable improvement compared to the IST EoS results than in 3-dimensional case. Remarkably,  it allows us  to extend the description of two-component SHDM EoS (\ref{Z mult Santos}) up to $\eta = 0.7$, i.e. for the whole gaseous phase with $\sqrt{\tilde \chi^2} \simeq 0.7$  percent  (see  the case $\alpha_1 = 1.076$  and $\alpha_2 = 1.575$ in Fig.  \ref{ISCT HD twocomp} and in Table \ref{HD_table}. 

\begin{table}
	\centering
	\caption{The best fit parameters of the ISCT EoS which reproduce the SHDM EoS on two intervals. The upper row corresponds to 
	to the fitting interval $\eta \in [0.;  0.7]$, while the  middle one provides an excellent description at the interval $\eta \in [0.;  0.5]$.
	The lowest row corresponds to the IST EoS with the case $\alpha_2 \neq \alpha_1$.}
	\label{HD_table}
	\begin{tabular}{l|l|l|l|l|l|l|l}	
		$\alpha_1$ & $\alpha_2$ & $\beta_1$ & $\beta_2$ & A & B & $\tilde \chi^2$ & $\eta$ \\ \hline
		\hline
		1.076 & 1.575 & 1.587 & 1.392 & 0.176 & 0.276 & 0.05 & 0 - 0.7 \\ \hline
		1.557 & 1.157 & 3.009 & 3.008 & 0.504 & -0.0099 & 0.008 & 0 - 0.5 \\ \hline
		4.736 & 1.539 &  &  &  &  & 4.48 & 0 - 0.5 \\
		
	\end{tabular}
\end{table}


\section{Discussion of results and perspectives}
\label{sec:discussing results}

In this work we analyzed the excluded volume of the multicomponent mixtures of hard spheres and hard discs.
With the help  of  the self-consistent approximation we derived the system of equations for the IST and ISCT EoS
which define the surface tension coefficient generated by the hard-core repulsion for the IST and ISCT  EoS
and the induced curvature tension coefficient for the ISCT case. 
In order to extend the applicability range of the IST  approach one has to include into 
the treatment the curvature tension coefficient. This is done in a general way by requiring that the relative importance  of surface and curvature tensions compared to the pressure should weaken at high packing fractions.
In practice, we suggest modifying the distribution functions entering   the expressions for  the induced
surface and curvature tension coefficients  in order to account for higher virial coefficients 
of the multicomponent  mixtures while keeping the correct values of the second virial coefficients. 

A comparison of the obtained ISCT EoS  with the well-known one-component  Carnahan-Starling EoS  for hard spheres and
Barrio-Solana  EoS for hard discs  showed us that  it is possible to accurately describe the compressibility factors of these equations for $\eta \le 0.3$ 
and $\eta\le 0.5$, respectively.  Basically the same results  are  found for the  two-component EoS of hard spheres and hard discs, if one 
describes them by the IST EoS. However, we found that the ISCT EoS is able to essentially improve the description of one- and two-component gases  of hard spheres and hard discs and to extend it to  the  entire gaseous phase with the accuracy  which exceeds the experimental  abilities of 
modern nuclear physics of intermediate and high energies. 
 
The fact that the same system of equations with the same number of phenomenological  parameters, i.e. the ISCT EoS, is able to accurately reproduce the well-known EoS of one- and two-component gases of hard spheres and hard discs  evidences that  the concept of ISCT 
  captures the correct physics in the grand canonical ensemble.  The great advantage of the IST and ISCT EoS for multicomponent mixtures compared
  to the other multicomponent EoS   is that the number of equations which are necessary to be solved does not depend on the number of different hard-core radii in the considered system. Therefore, the IST  and ISCT  can be used to formulate the  more realistic exactly solvable models  with the first order liquid-gas phase transition  in ordinary liquids, in nuclear liquid and in hadronic matter in order to improve the existing approaches (see, for instance, Refs. \cite{IST1, Reuter08, NewFisher1, NewFisher2, SMM1} and references therein).   As it was mentioned above the ISCT EoS  can also  be   used to improve the model of surface deformations of physical clusters  and to find out their surface entropy in a spirit of exactly solvable models of Refs. \cite{HDM1, HDM2}. 

Based on the mathematical  similarity between  the IH formula for  the second virial coefficient of convex hard particles and 
the expression  for the 
excluded volume of  two-component mixture  obtained here  we argue that after some modification the ISCT EoS can be used to model the 
multicomponent mixtures of convex hard particles of different shapes and sizes. Therefore, this direction of research may be useful 
 to many practical applications for which the exact treatment of EoS is mathematically too complicated, but comparatively   simple  numerical analysis like the molecular dynamics  can 
provide us with a few virial coefficients of multicomponent mixtures. 
{Clearly, the grand canonical  treatment of  such  multicomponent mixtures will be a good addition and alternative  to the well known
EoS of  multicomponent mixtures of convex hard particles of different shapes and sizes obtained in the canonical ensemble in Refs. \cite{Carnahan0,Carnahan1, Carnahan2}.}
Also, it is apparent  that the mathematical scheme outlined here can be generalized  to take into account for the attraction between the constituents. It is, however, clear that at low packing fractions such an attraction will decrease the value of  induced surface curvature coefficient, while  its  effect  on  the curvature tension coefficient  may depend on the attractive potential.  

The recent analysis of the quantum hard spheres \cite{QSTAT2019} shows that the set of equations which are found  directly from 
the quantum partition function  of a multicomponent mixture of  hard spheres  leads, in general,  to a different (and essentially more complicated) set of equations for the quantities 
$\overline{R}$ and
$\overline{R^2}$ than the one derived here. Nevertheless, in Ref.  \cite{QSTAT2019} it is shown that 
at low densities  the quantum equations  coincide  with the ones derived  here for  the induced surface tension  coefficient $\Sigma \equiv A \overline{R} p $ and for the curvature tension one  $ K\equiv  B \overline{R^2} p $.  Furthermore, as it is argued in  \cite{QSTAT2019}  the multicomponent  quantum  VdW  EoS with the  hard-core repulsion   is so complicated that  instead of its extrapolation to high densities it is more instructive   to extrapolate the quantum  analog of the system (\ref{Eq33n})-(\ref{Eq35n}) to high densities. 
Therefore, the equations for the induced surface tension  coefficient $\Sigma \equiv A \overline{R} p $ and for the curvature tension one  $ K\equiv  B \overline{R^2} p $ derived here for  the multicomponent VdW  EoS with the  hard-core repulsion  may be also important for formulating the realistic quantum EoS of  multicomponent   mixtures of hard discs, hard spheres and hard hyper-spheres of higher spatial dimensions. 

\vspace*{4.4mm}
\noindent
{\bf Acknowledgments.} 
The authors are thankful to 
B. E. Grinyuk,   O. I. Ivanytskyi, V.V. Sagun, S. N. Nedelko, E. G. Nikonov  and G. M. Zinovjev for fruitful discussions and valuable comments. 
The work of K.A.B. was supported by the Program of Fundamental Research in High Energy and Nuclear Physics launched by the Section of Nuclear Physics of the National Academy of Sciences of Ukraine.
The work of L.V.B. and E.E.Z. was supported by the Norwegian Research Council (NFR) under grant No. 255253/F53 CERN Heavy Ion Theory.  N.S.Ya., L.V.B. and K.A.B. thank the Norwegian Agency for International Cooperation and Quality Enhancement in Higher Education for the financial support under grant CPEA-LT-2016/10094 ``From Strong Interacting Matter to Dark Matter''. K.A.B. is also  grateful to the COST Action CA15213 ``THOR'' for supporting his networking.

\vspace*{0.22cm}

\appendix
\section*{Appendix: Expressions for particle density}

\section{Formulae for IST EoS}
\label{sec:appendix A}

In order to compare the IST EoS with the  Mansoori-Carnahan-Starling-Leland EoS \cite{MCSL} we need to have the explicit 
expressions for the particle number density of the $k$-th sort of particles. Hence, we consider the  system (\ref{eq for pressure}) and (\ref{eq for surf tension}), i.e., the partial pressure  $p_k$ and the partial surface-tension coefficient $\Sigma_k$ are 
defined as
\begin{eqnarray}
\label{EqIA}
p_k &=& T \phi_k \exp \left[ \frac{\mu_k}{T} - v_k \frac{p}{T} - s_k \frac{\Sigma}{T} \right]
\,, \\
\label{EqIIA}
\Sigma_k &=&  T  R_k \phi_k \exp \left[ \frac{\mu_k}{T} - v_k \frac{p}{T} - s_k \alpha_k 
\frac{\Sigma}{T} \right]  ~\quad
\\
&\equiv&  p_k R_k  \exp\left[ - s_k  (\alpha_k-1) \frac{\Sigma}{T}   \right] \,.
\end{eqnarray}
Then the total pressure and the total surface tension 
coefficient are defined as $p = \sum_k p_k$ and  $\Sigma = \sum_k \Sigma_k$, respectively. In  the 3-dimensional case (hard spheres) $v_k=\frac{4}{3}\pi R_k^3$ and $s_k = 4 \pi R_k^2$, while  in the 2-dimensional case (hard discs) $v_k= \pi R_k^2$ and $s_k = 2 \pi R_k$. 

Differentiating $p$  and $\Sigma$   with respect to the full chemical potential $\mu_k$ of  the hadron of sort $k$  one finds 
\begin{equation}
	\begin{pmatrix}
	a_{11} & a_{12} \\
	a_{21} & a_{22}
	\end{pmatrix}
	\cdot
	\begin{pmatrix}
	\frac{\partial p}{\partial \mu_k}  \\
	\frac{\partial \Sigma}{\partial \mu_k}
	\end{pmatrix} =
	\begin{pmatrix}
	\frac{p_k}{T} \\
	\frac{\Sigma_k}{T}
	\end{pmatrix}
\end{equation}
The coefficients $a_{kl}$ can be expressed in terms of  the partial pressures $\{ p_k\}$ and the 
partial surface tension coefficients $\{\Sigma_k\}$ as
\begin{eqnarray}
a_{11} = 1 +  \sum_k  v_k \frac{p_k}{T} \,,  &&  a_{12} = \sum_k s_k \frac{p_k}{T} \, , \\
a_{21} = \sum_k v_k \frac{\Sigma_k}{T} \, , &&  a_{22} = 1 +   \sum_k  s_k \alpha_k\frac{\Sigma_k}{T} \,.\quad 
\end{eqnarray}
Then the particle number density of  $k$-th sort of particle  is  given by
\begin{equation}\label{EqVIIIA}
\rho_k \equiv \frac{\partial  p}{\partial \mu_k} = \frac{1}{T} \cdot \frac{p_k \, a_{22} 
- \Sigma_k \, a_{12}}{a_{11}\, a_{22} - a_{12}\, a_{21} } \,.
\end{equation}

\section{Useful formulae for ISCT EoS}
\label{sec:appendix B}

Similarly to the IST EoS,  one  can calculate particle number densities for  ISCT EoS. The corresponding  coefficients are expressed in terms of partial  quantities $\{p_k\}$, $\{\Sigma_k\}$ and $\{K_k\}$ as
\begin{align*}
a &= 1 + \sum_k \alpha_k s_k \frac{\Sigma_k}{T} \,, & b &= \sum_k v_k \frac{\Sigma_k}{T} \,, & c &= \sum_k c_k \frac{\Sigma_k}{T} \,,\\
d &= 1 + \sum_k  \beta_k c_k \frac{K_k}{T} \,, & e &= \sum_k v_k \frac{K_k}{T} \,, & f &= \sum_k \alpha_k s_k \frac{K_k}{T} \,,\\
g &= 1 + \sum_k v_k \frac{p_k}{T} \,, & h &= \sum_k s_k \frac{p_k}{T} \,, & j &= \sum_k c_k \frac{p_k}{T} \,.
\end{align*}
Then for the  ISCT EoS the  particle number density of the $k$-th sort of particles  is given by
\begin{equation}
\hspace*{-2.mm}\rho_k = \frac{1}{T} \frac{(a d - c f) p_k - (d h - f j)\Sigma_k + (c h - a j) K_k}{g (a d - c f) - b (d h - f j) + e (c h - a j)} \,.
\end{equation}

\vspace*{2.2mm}
In  the 3-dimensional case (hard spheres) $v_k=\frac{4}{3}\pi R_k^3$, $s_k = 4 \pi R_k^2$ and $c_k = 4 \pi R_k$, while  in the 2-dimensional case (hard discs) $v_k= \pi R_k^2$, $s_k = 2 \pi R_k$ and $c_k = 2 \pi$.

	\end{document}